\documentclass[letterpaper]{article} % DO NOT CHANGE THIS
\usepackage{aaai24}  % DO NOT CHANGE THIS
\usepackage{times}  % DO NOT CHANGE THIS
\usepackage{helvet}  % DO NOT CHANGE THIS
\usepackage{courier}  % DO NOT CHANGE THIS
\usepackage[hyphens]{url}  % DO NOT CHANGE THIS
\usepackage{graphicx} % DO NOT CHANGE THIS
\usepackage{natbib}  % DO NOT CHANGE THIS AND DO NOT ADD ANY OPTIONS TO IT
\usepackage{caption} % DO NOT CHANGE THIS AND DO NOT ADD ANY OPTIONS TO IT
\usepackage{algorithm}
\usepackage{algorithmic}
\usepackage{booktabs}
\usepackage{amsthm}

\usepackage{amsmath}
\usepackage{amssymb}
\usepackage{subfigure}
\usepackage{multirow}
\usepackage{color}

\newtheorem{problem}{Problem}

\usepackage{newfloat}
\usepackage{listings}
\DeclareCaptionStyle{ruled}{labelfont=normalfont,labelsep=colon,strut=off} % DO NOT CHANGE THIS
\floatstyle{ruled}
\newfloat{listing}{tb}{lst}{}
\floatname{listing}{Listing}
\title{Spectral-Based Graph Neural Networks for Complementary Item Recommendation}
\author {
    Haitong Luo\textsuperscript{\rm 1, \rm 2},
    Xuying Meng\textsuperscript{\rm 1, \rm 5},
    Suhang Wang\textsuperscript{\rm 3},
    Hanyun Cao\textsuperscript{\rm 1, \rm 2}\\
    Weiyao Zhang\textsuperscript{\rm 1},
    Yequan Wang\textsuperscript{\rm 4},
    Yujun Zhang\textsuperscript{\rm 1, \rm 2, \rm 6}\thanks{Corresponding Author}
}
\affiliations {
    \textsuperscript{\rm 1}Institute of Computing Technology, Chinese Academy of Sciences,\\
    \textsuperscript{\rm 2}University of Chinese Academy of Sciences,\\
    \textsuperscript{\rm 3}Pennsylvania State University,\\
    \textsuperscript{\rm 4}BAAI,\\
    \textsuperscript{\rm 5}Purple Mountain Laboratories,\\
    \textsuperscript{\rm 6}Nanjing Institute of InforSuperBahn\\
    \{luohaitong21s, nrcyujun\}@ict.ac.cn
}

\usepackage{bibentry}
\begin{document}

\maketitle

\begin{abstract}
 % Modeling relationships between items is an important task in recommendation systems. The most common relationships we need to model are substitute and complement, in which for a query item, we need to recommend its substitutable or complementary items. Compared to substitutable relationships, modeling complementary ones is more challenging because complementary items are relevant but different while substitutable ones are always similar. 
Modeling complementary relationships greatly helps recommender systems to accurately and promptly recommend the subsequent items when one item is purchased. Unlike traditional similar relationships, items with complementary relationships may be purchased successively (such as iPhone and Airpods Pro), and they not only share relevance but also exhibit dissimilarity. Since the two attributes are opposites, modeling complementary relationships is challenging. Previous attempts to exploit these relationships have either ignored or oversimplified the dissimilarity attribute, resulting in ineffective modeling and an inability to balance the two attributes. Since Graph Neural Networks (GNNs) can capture the relevance and dissimilarity between nodes in the spectral domain, we can leverage spectral-based GNNs to effectively understand and model complementary relationships. 
In this study, we present a novel approach called Spectral-based Complementary Graph Neural Networks (SComGNN) that utilizes the spectral properties of complementary item graphs. We make the first observation that complementary relationships consist of low-frequency and mid-frequency components, corresponding to the relevance and dissimilarity attributes, respectively. Based on this spectral observation, we design spectral graph convolutional networks with low-pass and mid-pass filters to capture the low-frequency and mid-frequency components. Additionally, we propose a two-stage attention mechanism to adaptively integrate and balance the two attributes. Experimental results on four e-commerce datasets demonstrate the effectiveness of our model, with SComGNN significantly outperforming existing baseline models.

 % Modeling relationships between items is an important task in recommendation systems. Among these relationships, complementary ones are important, which means items with complementary functions are purchased together. Items with complementary relationships are not exactly similar, they also have different properties such as different functions, different appearances, and so on. Therefore, how to recommend complementary items for a given query item to users has become a non-trivial task in recommendation systems. The key to complementary relationships is relevance and dissimilarity, however, existing works do not deeply understand and effectively utilize the two attributes. This fact motivates us to propose Spectral-based Complementary Graph Neural Networks (SComGNN) to explore the spectral properties of complementary item graphs and conclude that \textbf{in the spectral domain, the complementary relationship is composed of low-frequency and mid-frequency components, which corresponds to the relevance and dissimilarity.}
 % Based on this spectral characteristic, we design different graph convolutional neural networks to obtain the low-frequency and mid-frequency components and propose a two-stage attention mechanism to adaptively integrate the two attributes. Experiments on four e-commerce datasets show the effectiveness of our model.
 
\end{abstract}

\section{Introduction}
%~\suhang{cite}
Complementary item recommendation \cite{liu2020decoupled, hao2020p, bibas2023semi} aims to suggest related items to users after they make a purchase in order to stimulate further purchases. To ensure the success of an e-commerce platform, it is crucial to model the complementary relationships between items. Complementary relationships involve items that are relevant yet dissimilar, as they are purchased together but serve different functions (e.g., iPhone and AirPods Pro). These properties make complementary relationships more challenging to model than traditional similarity relationships (also known as substitutable relationships). In this study, we focus on complementary item recommendation, i.e., given a query item, the goal is to recommend relevant yet dissimilar items to satisfy users' needs and encourage joint purchases.
% Among the various relationships between items, the most significant ones are substitutable and complementary relationships. Substitutable relationships refer to items that are similar, and users typically only purchase one of them (e.g., Apple and Samsung smartphones). Complementary relationships, on the other hand, involve items that are both relevant and different, as they are purchased together but serve different functions (e.g., Apple smartphones and Apple phone cases). 

 \begin{figure}
    \centering
    \includegraphics[scale=0.38]{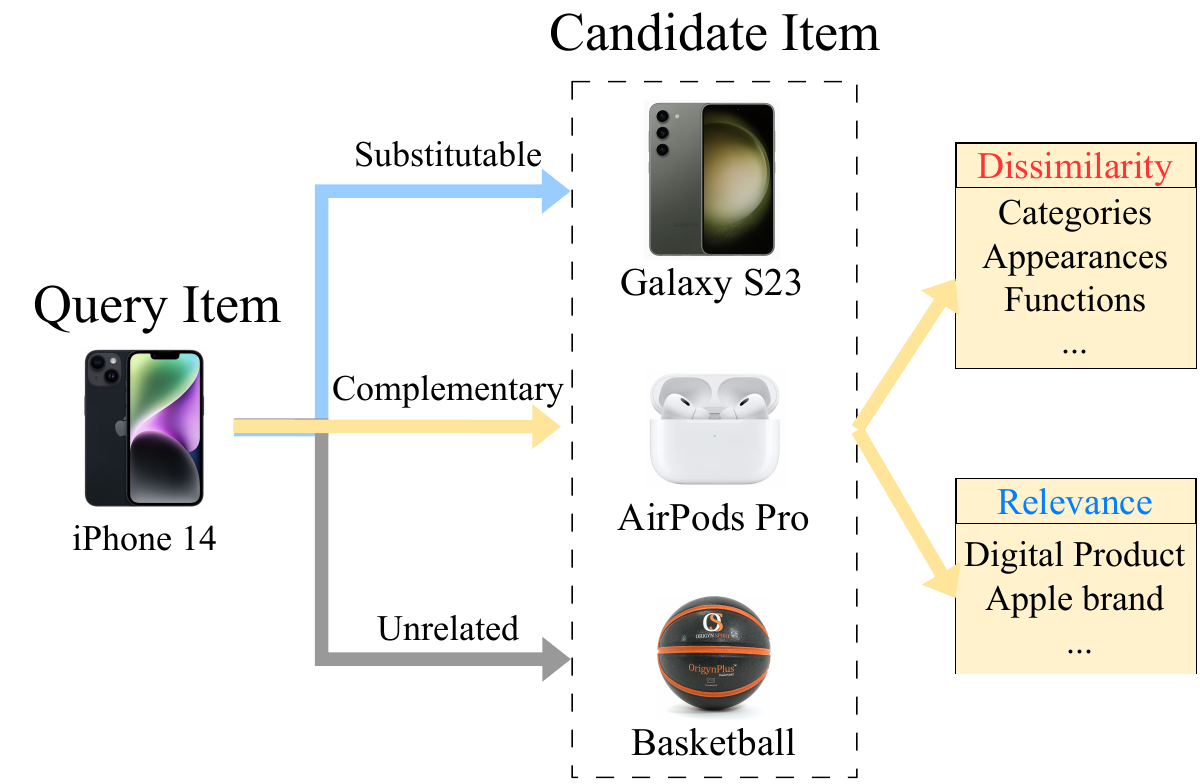}
    %\vspace{-5pt}
    \caption{Item relationships in recommender systems.}
    %\vspace{-15pt}
    \label{fig:illu}
\end{figure}
% Modeling complementary relationships between items is more challenging compared to substitutable relationships due to their specificity. When modeling substitutable relationships between items, it can be treated as a similarity evaluation problem. However, the same approach cannot be applied to modeling complementary relationships, as items with complementary relationships are both relevant and different from each other. Therefore, modeling complementary relationships between items is more challenging and requires further exploration. 

The core attributes of complementary relationships are relevance and dissimilarity. As shown in Figure \ref{fig:illu}, for iPhone and AirPods Pro, they are relevant as digital products under the Apple brand. Their dissimilarity lies in that they are different products with different functions and appearances. When recommending complementary products for users, it is crucial to understand and balance these two characteristics. On one hand, if we emphasize their relevance too much, it may lead to substitutable item recommendations. On the other hand, if we emphasize their dissimilarity too much, it may lead to recommendations for unrelated items.

Hence, researchers have made many efforts on complementary item recommendations. Some works \cite{mcauley2015inferring, wang2018path, cen2019representation, liu2020decoupled, chen2023enhanced} tentatively decouple and focus on complementary relations from item relationships. However, they ignore the dissimilarity attribute and only consider the relevance. 
To further model the dissimilarity attribute, recent works \cite{hao2020p, bibas2023semi} model the dissimilarity with category mapping networks that consider category diversity. However, these works still simplify the complementary relationships since the dissimilarity is not limited to categories. Without a deep understanding of these two attributes, existing works fail to model the essence of complementary relationships, which also leads to an inability to explore the trade-off between the two properties.
%P-Companion \cite{hao2020p} uses a category-transition autoencoder to satisfy the dissimilarity of complementary relationships. ALCIR \cite{bibas2023semi} uses category translators to map the category of complementary items and further satisfy the category diversity. 
% However, these works simplify the complementary item recommendations by just including item category diversity when modeling item relevance. They ignore the spectral properties of complimentary item graphs, and fail to model the essence of complementary relationships. Without a deep understanding of these two characteristics, it also leads to an inability to balance them effectively.
%~\suhang{cite}

Recent advances show that GNNs can capture the relevance and dissimilarities of nodes from the spectral domain \cite{wu2022beyond, tang2022rethinking}, which provides a promising direction to model the complementary relationships for simultaneously capturing the relevance and dissimilarity. Thus, in this work, we model complementary relationships with spectral-based GNNs. However, we are faced with two challenges: (1) \textbf{the lack of a deep understanding of complementary relationships from a spectral perspective}. Existing spectral-based GNNs do not explore and adapt to the spectral properties of complementary relationships, resulting in a gap between their spectral properties and the two attributes; (2) \textbf{the trade-off between the relevance and dissimilarity attributes}. Since the two attributes are opposites, over-emphasizing either one can lead to inaccurate complementary item recommendations. Therefore, it is crucial to strike a balance between these two attributes.

% \suhang{Talk about two challenges of adopting spectral-based GNN for complementary relationships. However, we are faced with two challenges: (1) the lack of a deep understanding of complementary relationships from a spectral perspective. Existing spectral-based GNNs are not designed for complementary item networks. It lacks a deep understanding of xxx; (ii) the trade-off between the relevance and dissimilarity attributes of complementary relationships. how to xxx} e.g., However, we are faced with two challenges: (i) Existing spectral-based GNNs are not designed for complementary product networks. It lacks a deep understanding of xxx; (ii) how to xxx}.
%In particualr, We aim to answer the following question---\textit{how can we understand the relevance and dissimilarity attributes of complementary relationships and how to balance them}. 
In an attempt to address these challenges, we first analyze complementary relationships from a spectral perspective on graphs and observe that the spectrum of the complementary item graph is mainly composed of low-frequency and mid-frequency components, which correspond to the relevance and dissimilarity characteristics respectively. Based on the observation, we design low-pass and mid-pass graph convolutional networks to decouple and extract the corresponding low-frequency relevance and mid-frequency dissimilarity components. To balance the two attributes, we propose a two-step attention mechanism to adaptively integrate and balance them. Our contributions are summarized as follows:
\begin{itemize}
    \item We conduct the first study to gain an understanding of the spectral properties of complementarity relationships based on GNNs, which associate the low-frequency and mid-frequency components with relevance and dissimilarity, respectively.
    \item We design a novel model with spectral-based GNNs and a two-stage attention mechanism, to decouple, extract and adaptively balance the low-frequency relevance and mid-frequency dissimilarity.
    \item We demonstrate the effectiveness of our proposed framework on four publicly available datasets, which outperforms the state-of-art approaches by a margin.
\end{itemize}

\section{Related Work}
In this section, we introduce related work on graph neural networks and complementary item recommendations.
\subsection{Graph Neural Networks} 

GNNs \cite{wu2020comprehensive} have shown great ability in modeling graph-structured data. Generally, GNNs can be classified into two main forms, i.e., spatial-based and spectral-based ones. \textit{Spatial-based GNNs} \cite{hamilton2017inductive, velivckovic2017graph, gao2018large, zhu2023you} operate in the spatial domain, where the graph convolution is defined in terms of the neighborhood structure of each node. 
%For example, GraphSage \cite{hamilton2017inductive} aggregates based on the nodes’ ego features and the fixed number of neighborhood features. Most recently, GSM-GNN \cite{zhu2023you} adaptively combines local and global information to enhance performance. 
\textit{Spectral-based GNNs} \cite{bruna2013spectral, defferrard2016convolutional, kipf2016semi, balcilar2020bridging, wu2022beyond} operate in the spectral domain, where the graph convolution filter is defined in terms of the eigenvectors of the graph Laplacian matrix. 
%Spectral CNN \cite{bruna2013spectral} firstly proposes a graph convolutional neural network in the spectral domain. Based on this, ChebyNet \cite{defferrard2016convolutional} and GCN \cite{kipf2016semi} approximate the spectral filters with different methods. 
Since GCN only utilizes low-frequency information \cite{balcilar2020bridging}, to broaden the available frequency bandwidth, recent studies \cite{balcilar2020bridging, wu2022beyond} attempt to designs filter functions to incorporate all the bands of graph signals.

\subsection{Complementary Item Recommendations}
To maximize profit and provide convenience to users, 
% E-commerce platforms usually recommend related items to customers during product searches. Hence, 
modeling item relationships is a crucial task in recommendation systems. However, existing works \cite{wang2011utilizing, yao2018judging,MengWSLCLZ18} often oversimplify item relationships as merely being ``related'', disregarding the fact that these relationships can be further categorized as substitutable or complementary ones. \textit{Complementary} relationships involve items that are relevant yet dissimilar, making the modeling of such relationships more challenging while \textit{substitutable} items are almost similar. To tackle this, one straightforward method is frequent pattern mining and association rules \cite{han2007frequent}.

Recently, deep learning methods have been applied to recommend complementary items. Some studies \cite{mcauley2015inferring, wang2018path, cen2019representation, liu2020decoupled, wu2022towards, chen2023enhanced} decouple and focus specifically on complementary relationships from item relationships to provide more precise recommendations. 
%For example, Sceptre \cite{mcauley2015inferring} adopts a topic modeling module to classify substitute and complementary relationships. PMSC \cite{wang2018path} introduces relation-aware parameters to model multiple relationships. GATNE \cite{cen2019representation} aims at modeling attributed multiplex networks to infer substitutable and complementary items effectively. DecGCN \cite{liu2020decoupled} works on decoupling and modeling substitutable and complementary relationships, enabling mutual benefits. LogiRec \cite{wu2022towards} introduces a logical reasoning network to solve the high-order complementary recommendation problem \cite{wu2022towards}. 
%While the above approaches initially focus on complementary relationships, 
However, they tend to ignore the dissimilarity attribute. In response, some works try to consider the dissimilarity attribute. For example, P-companion \cite{hao2020p} and ALCIR \cite{bibas2023semi} propose category mapping networks to recommend complementary items and include category diversity.
%Encore \cite{zhang2018quality} combines multiple sources of complement evidence, such as stylistic complements, to uncover high-quality complementary recommendations. 
% P-companion \cite{hao2020p} incorporates a type transition component for recommending complementary items to include category diversity. ALCIR \cite{bibas2023semi} proposes a semi-supervised approach that utilizes a category translator network to further focus on category diversity.

Since the dissimilarity is related to not only categories but also other features such as appearances and prices, these works still oversimplify the dissimilarity and do not have a deep understanding of the complementary relationships. Furthermore, their inability to accurately model and decouple the two attributes prevents them from striking a balance between relevance and dissimilarity.

%\suhang{few sentences state the novelty and difference of our method with the aforementioned methods.}

% In recent years some studies utilize deep learning methods to achieve complementary items recommendation. Sceptre. LogiRec proposes a logical reasoning network that includes probabilistic embedding of products and logical reasoning operations to solve the high-order complementary recommendation problem. Encore combines multiple sources of complement evidence(e.g., stylistic complements and functional complements) into their framework for uncovering high-quality complementary recommendations.
%  In addition, due to the powerful modeling capability of graph neural networks (GNNs), some work has also started to apply GNNs to achieve complementary item recommendations. Cen et al aims at modeling attributed multiplex networks and shows the state-of-the-art performance for inferring substitutable and complementary items. DecGCN model item substitutability and complementarity in separated embedding spaces and make the modeling of each relationship benefit from the other.

\section{Problem Statement and Motivation}
% In this sprovide a preliminary study on spectral analysis of complementary item graphs. 
We first provide preliminaries and definitions of our graph-based complementary item recommendation problem. 
% Based on the definition, we analyze the complementary item graph in the spectral domain on real-world datasets, which paves us a way to design a framework to model the relevance and dissimilarity of complementary relationships using GNNs.

\subsection{Notations and Problem Definition}
\label{sec:preliminaey_study}

Let $\mathcal{G} = {\{\mathcal{V}, \mathcal{\mathbf{X}}, {\mathcal{E}}}\}$ denotes the complementary item graph, where $\mathcal{V}$ is the set of nodes $ {\{v_1,...,v_N\}}$ and each node is an item.  $\mathcal{E}=\{e_{ij}\}$ is the set of undirected edges. Feature matrix $\mathcal{\mathbf{X}}\in \mathbb{R}^{N\times d}$ is made of $d-$dimensional features of $N$ nodes. Let $\mathbf{A} \in \mathbb{R}^{N\times N}$ denotes the adjacency matrix. $\mathbf{A}_{ij}=1$ if $v_i$ and $v_j$ are complementary; otherwise,  $\mathbf{A}_{ij}=0$. Let $\mathbf{D} \in \mathbb{R}^{N\times N}$ be the diagonal degree matrix with $\mathbf{D}_{ii}=\sum_{j}\mathbf{A}_{ij}$. The normalized graph Laplacian matrix $\mathbf{L}=\mathbf{I}-\mathbf{D}^{-\frac{1}{2}}\mathbf{A}\mathbf{D}^{-\frac{1}{2}}$, where $\mathbf{I}$ is an identity matrix. 
%In the complementary item graphs, nodes $\mathcal{V}$ represent items and edges $\mathcal{E}$ represent the complementary relationships between items. Once two items $v_i$ and $v_j$ are complementary, they will be linked and there exists an edge $e_{ij}$. 
% Generally, graph-based complementary item recommendation can be treated as a link prediction problem, which can be formally defined as:
With these notations, we formally define the problem of graph-based complementary item recommendation as:
\begin{problem}
    For a complementary item graph $\mathcal{G} = {\{\mathcal{V}, \mathcal{\mathbf{X}}, {\mathcal{E}}}\}$, where nodes denote items and edges denote complementary relationships, we aim to predict the probability of an edge $e_{i,j}$ when given two items $v_i$ and $v_j$, and find items being complementary accordingly. 
\end{problem}

 %Treating graph-based complementary item recommendation as a  link prediction problem, we aim to utilize the spectral characteristics to predict the probability of an edge $e_{i,j}$ when given two items $v_i$ and $v_j$, and find items being complementary accordingly.  

\subsection{Observations on Real-world Datasets}
\begin{table}[]
\small
\centering
% \scalebox{1}{
\begin{tabular}{@{}c|cccc@{}}
\toprule
Dataset  & Appliances & Grocery & Toys    & Home    \\ \hline
\# items    & 804       & 38548      &24638   & 75514   \\
\# edges    & 8290      & 642884     &614730  & 776766 \\
$S_{high}$    & 0.3408  & 0.4034 & 0.3150 & 0.4169 \\ \bottomrule
\end{tabular}
% }
%\vskip -1em
\caption{Statistics and $S_{high}$ of four datasets.}
% \vskip -1em
\label{tab:s_high}
\end{table}

To observe the spectral properties of the complementary item graphs, we first introduce two metrics \cite{tang2022rethinking}, i.e., \textbf{spectrum} and \textbf{high-frequency area}.

(1) The spectrum visualizes frequency distribution in spectral domains. It is plotted using eigenvalues $\mathbf{\lambda}$ as the x-axis and the spectral energy as the y-axis. The eigenvalues ${\mathbf{\lambda}}=\{\lambda_1,\lambda_2,...,\lambda_N\}$ and the corresponding eigenvectors $\mathbf{U}=(\mathbf{u}_1,\mathbf{u}_2,...,\mathbf{u}_N)$ can be obtained by the decomposition of the normalized Laplacian matrix $\mathbf{L}$, where eigenvalues $\mathbf{\lambda}$ also denote the frequencies of the graph. The spectral energy $\hat{x}_k^2/\sum_{i=1}^N\hat{x}_i^2$ is based on the graph Fourier transform $\mathbf{\hat{x}}=(\hat{x}_{1},\hat{x}_{2},...,\hat{x}_{N})^T = \mathbf{U}^T\mathbf{x}$, where $\mathbf{x}=({x}_{1},{x}_{2},...,{x}_{N})^T\in\mathbb{R}^N$ denotes one dimension of features from $N$ nodes. Since $\mathbf{\lambda}$ ranges from 0 to 2, we define $\mathbf{\lambda}$ close to 2 as high frequencies, $\mathbf{\lambda}$ close to 0 as low frequencies, and $\mathbf{\lambda}$ close to 1 as medium frequencies. Due to the huge computational effort of eigenvalue decomposition, only small-scale graph datasets can be drawn for spectrum plots.

\begin{figure}[t]
    \centering
    \includegraphics[scale=0.27]{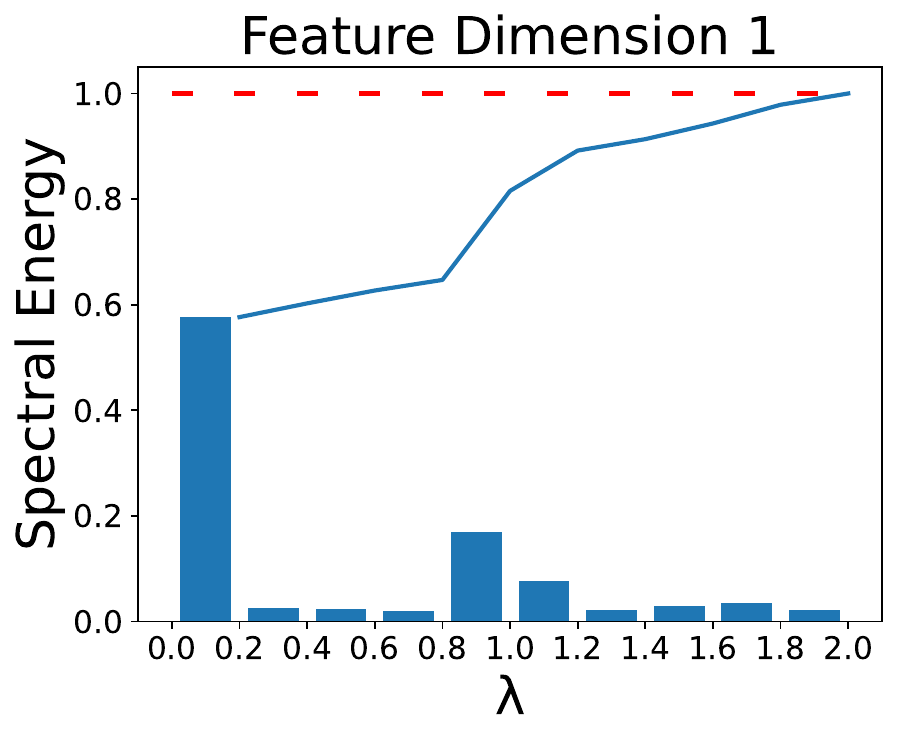}
    \includegraphics[scale=0.27]{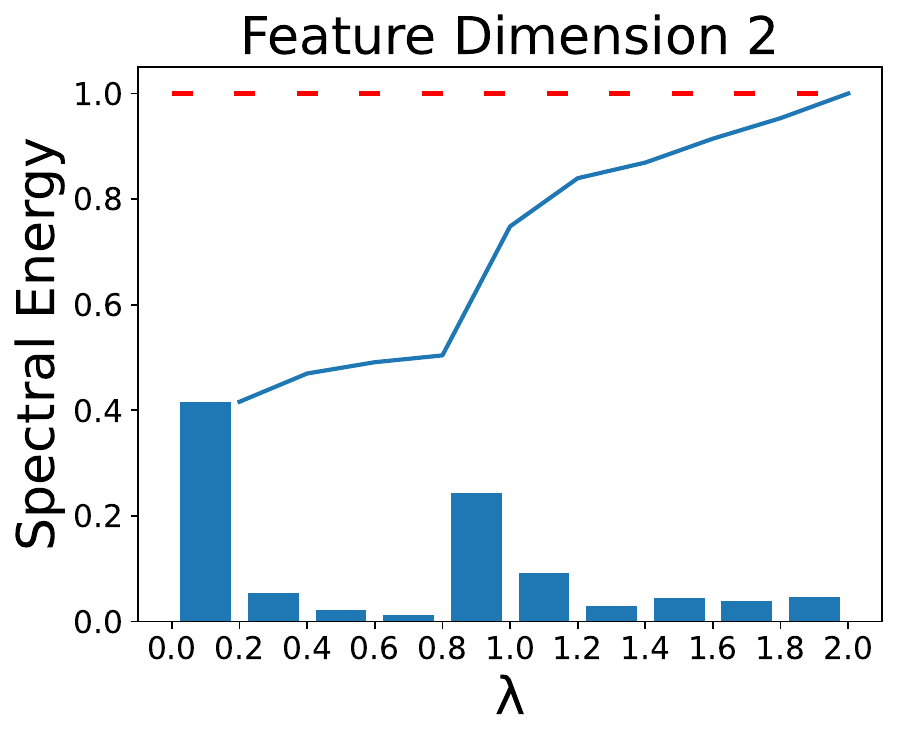}
    %\vskip -1em
    \caption{Spectral energy distribution of Appliances dataset.}
    %\vskip -1em
    \label{fig:spectrogram}
\end{figure}

(2) The high-frequency area $S_{high}$ denotes the area of the high-frequency region in the spectrum. It measures the area between the accumulated spectral energy curve (the solid lines in Figure \ref{fig:spectrogram}) and the curve with a y-value of 1 (the dashed lines in Figure \ref{fig:spectrogram}). Thus, $S_{high}$ is within $[0,2]$. Previous work \cite{tang2022rethinking} shows that $S_{high}$ can be obtained by $S_{high}=\frac{\sum_{k=1}^N\lambda_k\hat{x}_k^2}{\sum_{k=1}^N\hat{x}_k^2}=\frac{\mathbf{x}^T\mathbf{L}\mathbf{x}}{\mathbf{x}^T\mathbf{x}}$. There is no need for eigenvalue decomposition, making it feasible for large-scale datasets. The larger the $S_{high}$ is, the more the mid- and high-frequency components are.

%For $\mathbf{x}=(x_1,x_2,...,x_N)^T \in \mathbb{R}^N$, its graph Fourier transform is $\mathbf{\hat{x}}=(\hat{x}_1,\hat{x}_2,...,\hat{x}_N)^T = \mathbf{U}^T\mathbf{x}$. The eigenvalues $\mathbf{\lambda}$ are also the frequencies of the graph, and for $\lambda_k$, we can denote the \textit{spectral energy distribution} as $\hat{x}_k^2/\sum_{i=1}^N\hat{x}_i^2$. Considering that the maximum eigenvalue of normalized Laplacian matrix $\mathbf{L}$ is 2 and the minimum eigenvalue is 0, we classify eigenvalues near 2 as high frequencies, eigenvalues near 0 as low frequencies, and eigenvalues near 1 as medium frequencies. Based on this, we can plot the spectrum with the x-axis being the frequencies $\lambda$ and the y-axis being the \textit{spectral energy distribution}.

Based on the two metrics, we conduct our analysis on four real-world datasets obtained from Amazon \cite{he2016ups}, i.e., ``Appliances'', ``Grocery and Gourmet Food'' (abbreviated as Grocery), ``Toys and Games'' (Toys), and ``Home and Kitchen'' (Home). The details of the datasets can be found in the experiment section. ``Appliances'' is a small-scale dataset, while the others are large-scale datasets. In each dataset, nodes represent items and edges represent complementary relationships. 

We draw the spectrums for the small-scale dataset and compute high-frequency area $S_{high}$ for all datasets. Since both the spectrum and $S_{high}$ are obtained based on a single feature dimension, we randomly select two dimensions to plot the spectrum in Figure \ref{fig:spectrogram} and calculate the average value of all feature dimensions to obtain $S_{high}$ in Table \ref{tab:s_high}. In Figure \ref{fig:spectrogram}, the histogram denotes the spectral energy distribution and the solid curve denotes the accumulated spectral energy distribution.

From Figure \ref{fig:spectrogram} and Table \ref{tab:s_high}, we can conclude that the complementary item graph is mainly composed of low-frequency and mid-frequency components in the spectral domain. In detail, (1) from Figure \ref{fig:spectrogram}, the spectrum shows that $\lambda$ with low and medium values have larger spectral energies, which means the complementary relationship is composed of low-frequency and mid-frequency components; and (2) from Tabel \ref{tab:s_high}, the high-frequency area $S_{high}$ of all the datasets fall between 0 and 1, indicating that similar to the Appliances dataset, the spectrum of the other three datasets are mainly composed of low-frequency and mid-frequency parts. Additionally, as $S_{high}$ is below 0.5, the low-frequency component is greater than the mid-frequency component. 

Since the more similar a node is to its neighbors in the spatial domain, the lower the corresponding frequency component is in the spectral domain \cite{wu2022beyond, bo2021beyond}, we can regard the low-frequency component as the relevance attribute and the mid-frequency component as the dissimilarity attribute. To further verify it, we conduct a case study experiment in the experiment section. In this way, we bridge the gap between the properties of the complementary relationship and the spectral characteristics. 
 %Specifically: (1) The spectrum shows that the complementary relationship is composed of low-frequency and mid-frequency components, which correspond to the relevance and diversity of complementary relationships, respectively; (2) The high-frequency area indicates that, compared to the substitute relationship (which is mainly composed of low-frequency components), the complementary relationship has more mid-to-high frequency components; (3) From the analysis of three different features, for complementary relationships, category features have more high-frequency components, while brand features have fewer high-frequency components. This is because complementary relationships usually occur between different products of the same brand, so the products that have complementary relationships are usually similar in the brand.

 %Through spectrum analysis, we can conclude that the mid-frequency component of the complement graph is particularly important. However, existing methods have ignored this point. Therefore, making full use of the mid-frequency component of the complement graph can be very helpful in improving the performance of complementary item recommendations.

\begin{figure*}[]
    \centering
    \includegraphics[scale=0.35]{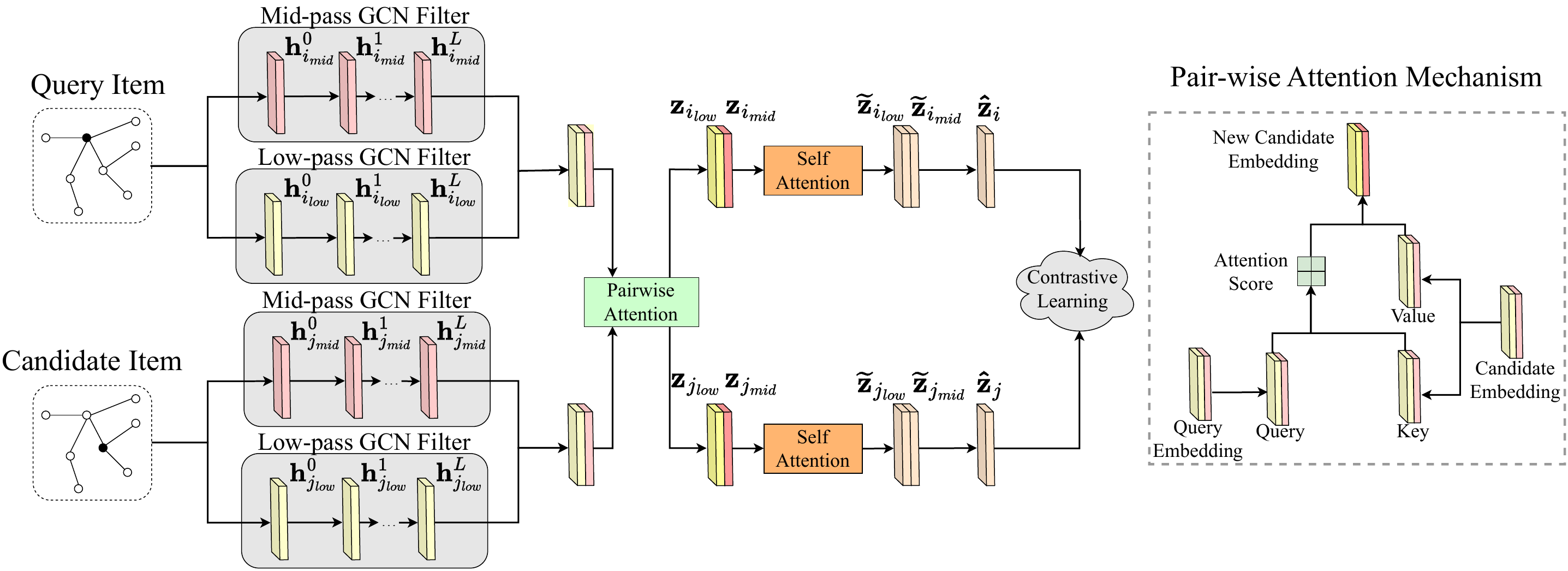}
    % \vskip -0.5em
    \caption{The overall framework of our proposed model SComGNN.}
    \label{fig:frame}
    %\vspace{-10pt}
\end{figure*}

\section{Methodology}
% In this section, we first overview our spectral-based complementary graph neural networks SComGNN. Then, the design of ScomGNN is introduced in detail, including the spectral-based GCN filters, the tow-stage attention mechanism, and the contrastive learning optimization.

% In this section, we introduce our proposed model. Section \ref{sec:overview} gives an overview of our proposed framework. Section \ref{sec:filter} presents the low-pass and mid-pass GNNs to filter out the relevance and dissimilarity of complementary relationships, separately. Then Section \ref{sec:attention} introduces the two-stage attention mechanism including pairwise attention and self-attention mechanism for the adaptive integration of low-frequency relevance and mid-frequency dissimilarity. Finally Section \ref{sec:loss} introduces the training and optimization procedure.

% \subsection{Model Overview}
\label{sec:overview}
% Based on the above observation, we correspond the relevance and dissimilarity attributes to the low-frequency and mid-frequency components in the spectral domain, respectively. Thus, we propose a novel framework to model the low-frequency and mid-frequency components in the spectral domain for complementary item prediction. An illustration of   the proposed framework is given in Figure \ref{fig:frame}, which is composed of three important components, i.e., the spectral-based GCN filters, the two-stage attention mechanism, and the contrastive learning optimization. (1) To model the low-frequency and mid-frequency components, we decouple and extract the low-frequency relevance and mid-frequency dissimilarity using specialized GCN filters. (2) After decomposing, how to integrate these two attributes presents a challenge. Since manually determining the importance of each attribute is challenging, we subsequently propose a two-stage attention mechanism to adaptively integrate them. In detail, we first employ a pairwise attention mechanism to let one item determine the significance of the relevance and dissimilarity of another and then utilize a self-attention mechanism to integrate the relevance and dissimilarity by itself. (3) Finally, we optimize our model using the contrastive learning approach. Next, we give the details of each component.
Based on our observations, we propose a novel framework for complementary item prediction by modeling the two attributes in the spectral domain. As illustrated in Figure \ref{fig:frame}, the framework consists of three key modules: spectral-based GCN filters, a two-stage attention mechanism, and contrastive learning optimization. (1) To model the low-frequency relevance and mid-frequency dissimilarity, we decouple and extract them using specialized GCN filters. (2) Integrating these attributes poses a challenge, as manually determining their importance is difficult. Thus we introduce a two-stage attention mechanism that adaptively integrates the attributes. It utilizes a pairwise attention mechanism to determine the significance of relevance and dissimilarity between by item pairs, followed by a self-attention mechanism that integrates the attributes independently. (3) Finally, we optimize our model using contrastive learning. In the following sections, we provide details of each module.

\subsection{Spectral-based GCN Filters}
\label{sec:filter}
To model the low- and mid- components of the complementary item graph, we decouple the low-frequency relevance and mid-frequency dissimilarity using specialized GCN filters. We will first introduce the unified form of spatial-based and spectral-based GCNs. Based on it, we then design spectral-based low-pass and mid-pass filters and turn them into spatial forms for implementation.

\subsubsection{Unified form of Spatial-based and Spectral-based GCNs}
GCNs can be explored from both spatial and spectral domain perspectives. Though the two approaches start from different domains, they can be interchanged \cite{balcilar2020bridging}. The GCN propagation can be formulated as:
\begin{equation}
\mathbf{H}^{l+1}=\sigma(\sum_{k=1}^{K}\mathbf{C}_{k}\mathbf{H}^{l}\mathbf{W}^{l}_{k}),
\end{equation}
where $\sigma$ is the activation function, $K$ is the number of filters, $\mathbf{H}^{l}$ denotes the node representation at layer $l$, and $\mathbf{W}^{l}_{k}$ is a learnable weight matrix of the filter $k$ at layer $l$. Here $\mathbf{C}_{k}$ is the graph convolution kernel in the spatial domain, which can be formulated in the spectral domain as:
\begin{equation}
    \mathbf{C}_{k}=\mathbf{U}\textit{diag}(F_{k}(\mathbf{\mathbf{\lambda}}))\mathbf{U}^T,
    \label{eq:spatial_gcn}
\end{equation}
where $\mathbf{U}$ and $\mathbf{\lambda}$ are the eigenvectors and the eigenvalues of the normalized graph Laplacian matrix $\mathbf{L}$. Here \textit{diag} represents the diagonal matrix with specified elements. $F_{k}(\mathbf{\lambda})$ is the graph convolutional filter in the spectral domain and is a function with $\mathbf{\lambda}$ as the independent variable. Eq. (\ref{eq:spatial_gcn}) can also be formulated as:%\suhang{explain diag} 
\begin{equation}
    F_{k}(\mathbf{\lambda}) = \textit{diag}^{-1}(\mathbf{U}^T\mathbf{C}_{k}\mathbf{U}).
    \label{eq:spectral_gcn}
\end{equation}
The key to the spatial-based GNNs is the design of $\mathbf{C}_{k}$, while the key to the spectral-based GNNs is the design of $F_{k}(\mathbf{\lambda})$. With Eq. (\ref{eq:spatial_gcn}) and Eq. (\ref{eq:spectral_gcn}), these two convolutional kernels can be converted to each other. Since spatial-based GNNs are generally easier to understand and implement than spectral-based ones, Eq. (\ref{eq:spatial_gcn}) inspires us to design the spectral convolutional kernel $F_{k}(\mathbf{\lambda})$ first, and then convert it to a spatial form to implement it, like what existing works do \cite{kipf2016semi, wu2022beyond}.

\subsubsection{Spectral-based Filters}
Existing spectral-based GNNs \cite{bruna2013spectral, defferrard2016convolutional, kipf2016semi} design different $F_{k}(\mathbf{\lambda})$ to obtain different GNN models. Since complementary item graphs are mainly composed of low and mid-frequency components in the spectral domain, our goal is to design a low-pass and a mid-pass GCN filter to extract the low and mid-frequency components, respectively, and filter out other components.
%Considering that the maximum eigenvalue of matrix $\mathbf{L}$ is 2 and the minimum eigenvalue is 0, we classify eigenvalues near 2 as high frequencies, eigenvalues near 0 as low frequencies, and eigenvalues near 1 as medium frequencies. 
%We design a low-pass and a mid-pass GCN filter to extract the two components respectively.
% Similar to \cite{wu2022beyond}, in order to avoid complex calculations, we design filters as simple as possible. 
 As the convolution of the spatial domain is equal to the product of the spectral domain, the larger the amplitude value, the more of the corresponding frequency component is retained. To avoid complex calculations, we design the spectral convolution kernel of the low-pass filter as a linear decreasing function of $\mathbf{\lambda}$:
\begin{equation}
    F_{low}(\mathbf{\lambda})=1-\mathbf{\lambda}/2.
\end{equation}
With Eq. (\ref{eq:spatial_gcn}), we can turn it into a spatial form:
\begin{equation}
    \mathbf{C}_{low}=(\mathbf{\widetilde{A}}+\mathbf{I})/2,
    \label{eq:low_pass}
\end{equation}
where $\mathbf{\widetilde{A}}=\mathbf{D}^{-1/2}\mathbf{A}\mathbf{D}^{-1/2}$, denoting the normalized adjacency matrix.
Linear functions can be implemented as high-pass or low-pass in increasing or decreasing form, however, it is not possible to implement a mid-pass filter, where the mid-frequency component is retained while others are filtered out. Therefore, we use the quadratic function of $\lambda$ to realize the spectral convolution kernel of the mid-pass filter:
\begin{equation}
    F_{mid}(\mathbf{\lambda})=-(\mathbf{\lambda}-1)^2+1.
\end{equation}
With Eq. (\ref{eq:spatial_gcn}), it can also be formulated in the spatial domain:
\begin{equation}
    \mathbf{C}_{mid}=\mathbf{I}-\mathbf{\widetilde{A}}^2.
    \label{eq:mid_pass}
\end{equation}

% To better understand the low-pass and mid-pass GCN filters, we plot the spectral convolutional kernels $F_{low}(\mathbf{\lambda})$ and $F_{mid}(\mathbf{\lambda})$ (i.e., the spectrums) and present toy examples of the corresponding spatial convolution kernels $\mathbf{C}_{low}$ and $\mathbf{C}_{mid}$ in Figure \ref{fig:example}. 

\begin{figure}[t]
    \centering
    \includegraphics[scale=0.25]{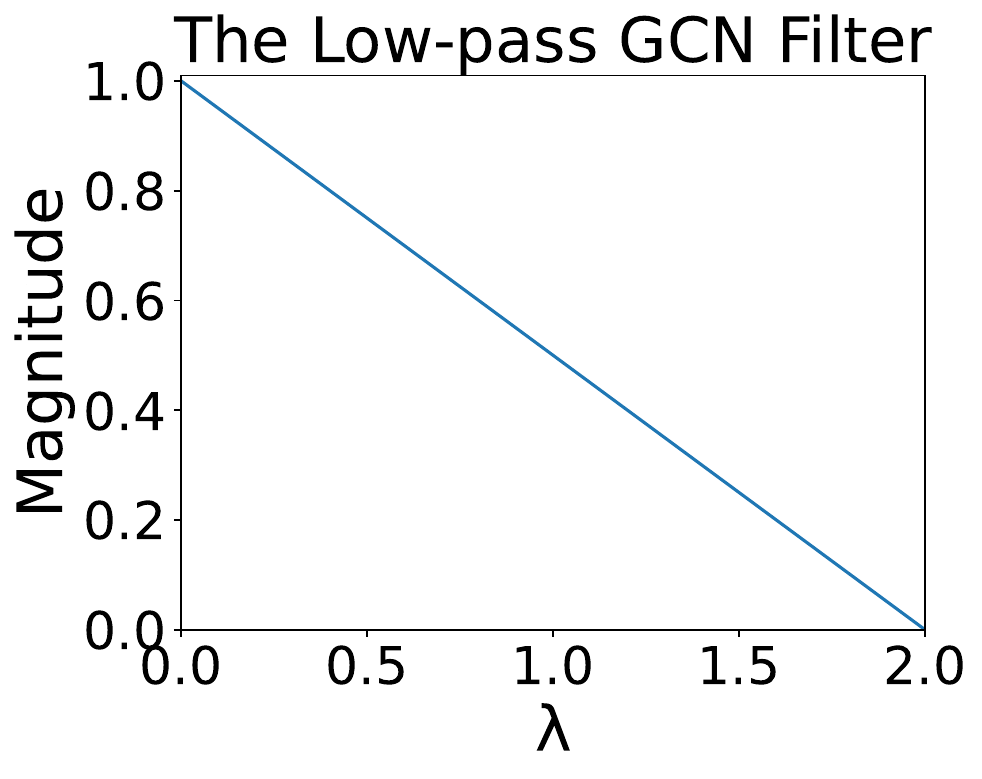}
    \includegraphics[scale=0.25]{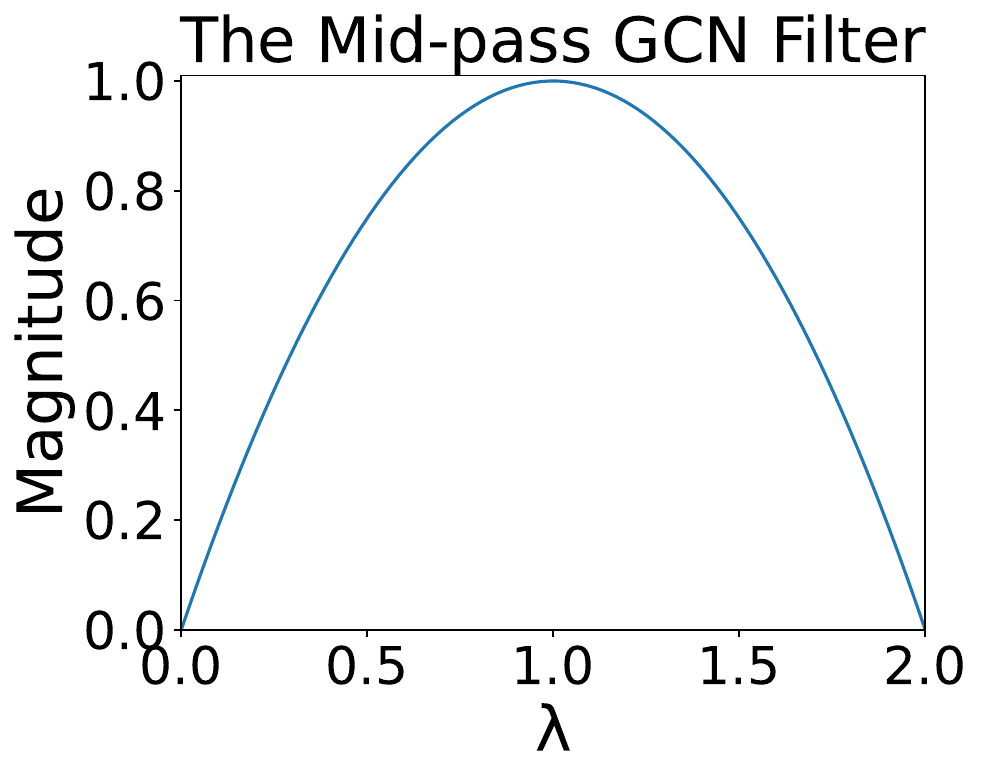}
    %\vskip -1em
    \caption{Spectrums of low-pass and mid-pass GCNs.}
    %\vskip -1em
    \label{fig:spectral_example}
    %\suhang{make the font larger, those notations are difficult to read}
\end{figure}

% In Figure \ref{fig:spectral_example}, the low-pass filter retains the low-frequency part and filters out other parts while the mid-pass filter retains the middle-frequency part and filters out other parts.  
% In Figure \ref{fig:spatial_example}, we present toy examples of the corresponding spatial convolutional kernels. Here node 1 is the central node thus we aim to aggregate the neighbors of node 1. With the low-pass filter, node 1 and its first-order neighbors undergo a summing operation, while with the mid-pass filter, node 1 subtracts information from its second-order neighbors. Therefore, in the spatial domain, the low-pass filter smoothes the self-information of the nodes and the information of the neighbors, while the middle-pass filter obtains the difference between the self-information of the nodes and the information of the second-hop neighbors. Thus we can learn that the low-pass filter extracts the relevance between nodes and neighbors, while the mid-pass filter extracts the dissimilarity between nodes and neighbors.
To better understand the low-pass and mid-pass filters, we look at both the spectral and spatial domains. For the spectral domain, we plot the spectral convolutional kernels $F_{low}(\mathbf{\lambda})$ and $F_{mid}(\mathbf{\lambda})$ (i.e., the spectrums) in Figure \ref{fig:spectral_example}. As shown in the spectrums, the low-pass filter retains the low-frequency part and filters out other parts while the mid-pass filter retains the middle-frequency part and filters out other parts. For the spatial domain, with Eq. (\ref{eq:low_pass}) and Eq. (\ref{eq:mid_pass}), our low-pass filter aggregates a node’s self-information with its neighborhood information while the mid-pass filter obtains the difference between a node’s self-information and its two-hop neighborhood information. Therefore, we can conclude that the low-pass filter extracts the relevance between nodes and neighbors, while the mid-pass filter extracts the dissimilarity between nodes and neighbors.

% \begin{figure}[t]
%     \centering
%     \subfigure[Spectrum of low-pass and mid-pass GCNs.]{
%     \includegraphics[scale=0.265]{AnonymousSubmission/LaTeX/pics/low-pass.pdf}
%     \includegraphics[scale=0.265]{AnonymousSubmission/LaTeX/pics/mid-pass.pdf}
%     \label{fig:spectral_example}
%     }
%     \vskip -0.1em
%     \subfigure[Toy examples of low-pass and mid-pass GCNs.]{
%     \includegraphics[scale=0.7]{AnonymousSubmission/LaTeX/pics/low_example_裁剪页面.pdf}
%     \includegraphics[scale=0.7]{AnonymousSubmission/LaTeX/pics/mid_example.pdf}
%     \label{fig:spatial_example}
%     }
%     \vskip -1em
%     \caption{Spectrum and toy examples of low-pass and mid-pass GCNs.}
%     % \vskip -1em
%     \label{fig:example}
% \end{figure}

With the low-pass and mid-pass GCN filters, we can extract the low and mid-frequency components from the complementary item graph, which correspond to the relevance and dissimilarity attributes, respectively. We formulate it as:
 \begin{gather}
    \mathbf{H}_{mid}^{l}=Relu(\mathbf{C}_{mid}\mathbf{H}_{mid}^{l-1}\mathbf{W}^{l-1}), \\
    \mathbf{H}_{low}^{l}=Relu(\mathbf{C}_{low}\mathbf{H}_{low}^{l-1}\mathbf{W}^{l-1}),
    \label{eq:spectral_filter}
 \end{gather}
where $\mathbf{H}_{low}^{l}$ and $\mathbf{H}_{mid}^{l}$ are the low-frequency and mid-frequency node representation matrix at layer $l$, $\mathbf{W}^{l-1}$ is the weight matrix. Note that, $\mathbf{H}_{low}^0=\mathbf{H}_{mid}^0=\mathbf{X}$.
%Due to the simple implementation in the spatial domain, the computational complexity of the low-pass and mid-pass GCN filters is $\mathcal{O}(|\mathcal{E}|)$ and $\mathcal{O}(2|\mathcal{E}|)$ respectively, where $|\mathcal{E}|$ is the number of edges. Therefore, the total time complexity of spectral-based GCN filters is $\mathcal{O}(3|\mathcal{E}|)$.
 \subsection{Tow-stage Attention Mechanism}  \label{sec:attention}
Since items with complementary relationships are relevant yet dissimilar, it is challenging to determine manually which attribute is more crucial when making complementary relationship predictions. Even for the same item, the significance of these two attributes may differ in different item pairs. 
% For instance, when making complementary relationship predictions for item pairs $(v_i, v_j)$ and $(v_i, v_k)$, the mid-frequency dissimilarity of item $v_i$ may be more crucial in pair $(v_i, v_j)$ while the low-frequency relevance of item $v_i$ may be more important in pair $(v_i, v_k)$. 
To tackle this problem, we propose a two-stage attention mechanism to merge the two attributes adaptively, which is composed of pairwise attention in item pairs and self-attention independently.
%\suhang{to xxx} and self-attention \suhang{to xxx}.
%\suhang{make it consistent in the paper, use $v_i$ and $v_j$} 

\subsubsection{Pairwise Attention Mechanism.}
   \begin{table*}[]
  \small
 \centering
 \scalebox{1}{
 \setlength{\tabcolsep}{1.1mm}{
 \begin{tabular}{c|c|ccc|ccc|ccc|ccc}
 \toprule
 \multirow{2}{*}{Method} & Datasets    & \multicolumn{3}{c|}{Appliances} & \multicolumn{3}{c|}{Toys} & \multicolumn{3}{c|}{Grocery} & \multicolumn{3}{c}{Home} \\ \cline{2-14} 
                        & Metric      & HR@5  & HR@10  & NDCG    & HR@5  & HR@10  & NDCG   & HR@5 & HR@10 & NDCG   & HR@5 & HR@10 & NDCG \\ \hline \hline
 \multirow{6}{*}{Baselines}               & GIN   & 0.4347   & 0.6226   & \underline{0.4279}      & 0.5242      & 0.7408        & \underline{0.4866}          & 0.4344     & 0.6107       & 0.4425   & 0.4843     & 0.6552       & 0.4681         \\
                        & GraghSage       &\underline{0.4402}    & \underline{0.6574}   &0.4215    & \underline{0.5313}      & \underline{0.7514}    & 0.4863   & \underline{0.6255}      &\underline{0.8000}      &\underline{0.5359}  & \underline{0.7272}     & \underline{0.8417}   & \underline{0.6061}     \\ \cline{2-14}
                        & Popularity  &0.2040  & 0.3208         & 0.2906        & 0.1809      & 0.2816        & 0.2914  & 0.2556  & 0.3690       & 0.3392  & 0.2263     & 0.3505  & 0.3223           \\
                        & DCF         & 0.3630       & 0.5366         & 0.3817     & 0.4876      & 0.6714        & 0.4661   &0.5991      &0.7574  & 0.5326  & 0.6846     & 0.7798      & 0.6015         \\
                        & P-Companion & 0.3545       & 0.5414         & 0.3759     & 0.4098      & 0.6017         & 0.3923    & 0.4943     & 0.6774   & 0.4152   & 0.5847   & 0.7220   & 0.5145         \\
                        & ALCIR       &  0.3754      &  0.5394     & 0.3792     & 0.3930      & 0.5959        & 0.3994   & 0.5067   & 0.6892  & 0.4614    & 0.5411     & 0.6885  & 0.4826    \\ \hline \hline
{Ours}  & SComGNN     & \textbf{0.4919}    & \textbf{0.7127}      & \textbf{0.4377}        & \textbf{0.6561}      & \textbf{0.8589}      & \textbf{0.5501}    & \textbf{0.7207}   & \textbf{0.8565}   & \textbf{0.5959}    & \textbf{0.7943}     & \textbf{0.8789}    & \textbf{0.6610}   \\ \bottomrule
 \end{tabular}}}
 %\vskip -0.5em
  \caption{Performance comparison on four datasets.}
\label{tab:overall_perform}
%\vspace{-5pt}
 \end{table*}
We first adopt a pairwise attention mechanism to integrate the low- and mid-frequency components of items in pairs. For example, given an item pair $(v_i, v_j)$, item $v_j$ determines the proportion of low-frequency relevance and mid-frequency dissimilarity of item $v_i$, and vice versa. 
With the low-pass and mid-pass GCN filters, we obtain the low and mid-frequency item representation matrix $\mathbf{H}_{low}^L$ and  $\mathbf{H}_{mid}^L$, where $L$ is the depth of propagation layers. For item $v_i$, we denote its embeddings as $\mathbf{h}_{i_f}^L$, where $f \in \{low,mid\}$. With these notations, we first introduce how the low-frequency component of item $v_j$ selects and integrates the two embeddings of item $v_i$: 
\begin{small}
\begin{gather}
   \mathbf{z}_{i_{low}}=\sum_{f}\alpha_{j_{low},i_{f}}\mathbf{{h}}_{i_f}^{L}, \quad
   \alpha_{j_{low},i_f}=\frac{\exp({\mathbf{h}_{j_{low}}^{L}}^T\mathbf{h}_{i_f}^{L})}{\sum_f\exp({\mathbf{h}_{j_{low}}^{L}}^T\mathbf{h}_{i_f}^{L})},
   \label{eq:pair_low}
\end{gather}
\end{small}
where $\mathbf{h}_{j_{low}}^L$ denotes the low-frequency embedding of item $v_j$. Here $\alpha_{j_{low},i_f}$ denotes the proportion of low-frequency and mid-frequency embeddings of item $v_i$. $\mathbf{z}_{i_{low}}$ denotes the integrated representation of item $v_i$ decided by low-frequency component of item $v_j$.

Also, the two embeddings of item $v_i$ can be selected and integrated by the mid-frequency component of item $v_j$:
\begin{small}
\begin{gather}
   \mathbf{z}_{i_{mid}}=\sum_{f}\alpha_{j_{mid},i_{f}}\mathbf{{h}}_{i_f}^{L}, \quad
   \alpha_{j_{mid},i_f}=\frac{\exp({\mathbf{h}_{j_{mid}}^{L}}^T\mathbf{h}_{i_f}^{L})}{\sum_f\exp({\mathbf{h}_{j_{mid}}^{L}}^T\mathbf{h}_{i_f}^{L})},
   \label{eq:pair_mid}
\end{gather}
\end{small}
where $\mathbf{h}_{j_{mid}}^L$ denotes the mid-frequency embedding of item $v_j$. Here $\alpha_{j_{low},i_f}$ denotes the proportion of two embeddings of item $v_i$. $\mathbf{z}_{i_{mid}}$ denotes the integrated representation of item $v_i$ decided by mid-frequency component of item $v_j$.

By the pairwise attention mechanism with item $v_j$, we obtain the integrated embedding $\mathbf{z}_{i_{low}}$ and $\mathbf{z}_{i_{mid}}$ of item $v_i$. Here the $low$ and $mid$ in the subscripts no longer denote the frequency component in item $v_i$, but rather which frequency component of item $v_j$ it is integrated from. For item $v_j$ in pair $(v_i, v_j)$, we can do the same pairwise attention step to obtain $\mathbf{z}_{j_{low}}$ and $\mathbf{z}_{j_{mid}}$.

\subsubsection{Self-attention Mechanism.}
% with the \textit{responded representation}, we utilize the self-attention step to let the item decide which part is more important by itself.
After the pairwise attention step, the low and mid-frequency components of items $v_i$ and $v_j$ have been selected and integrated by each other. Next, we use the self-attention mechanism to further adaptively integrate the low and mid-frequency components by themselves. Similar to the above pairwise attention step, two embeddings of item $v_i$ can be selected and integrated by $\mathbf{z}_{i_{low}}$:
\begin{small}
\begin{gather}
   \mathbf{\widetilde{z}}_{i_{low}}=\sum_{f}\beta_{i_{low},i_{f}}\mathbf{z}_{i_f}, \quad 
   \beta_{i_{low},i_f}=\frac{\exp({\mathbf{z}_{i_{low}}}^T\mathbf{z}_{i_f})}{\sum_f\exp({\mathbf{z}_{i_{low}}}^T\mathbf{z}_{i_f})},
   \label{eq:self_low}
\end{gather}
\end{small}
where $\beta_{i_{low},i_f}$ denotes the proportion of two embeddings of item $v_i$, $\mathbf{\widetilde{z}}_{i_{low}}$ denotes the further integrated representation of item $v_i$ by $\mathbf{z}_{i_{low}}$ after the self-attention step.
Also, the two embeddings of $v_i$ can be selected and integrated by $\mathbf{z}_{i_{mid}}$:
\begin{small}
\begin{gather}
   \mathbf{\widetilde{z}}_{i_{mid}}=\sum_{f}\beta_{i_{mid},i_{f}}\mathbf{z}_{i_f}, \quad
   \beta_{i_{mid},i_f}=\frac{\exp({\mathbf{z}_{i_{mid}}}^T\mathbf{z}_{i_f})}{\sum_f\exp({\mathbf{z}_{i_{mid}}}^T\mathbf{z}_{i_f})},
   \label{eq:self_mid}
\end{gather}
\end{small}
where $\beta_{i_{mid},i_f}$ denotes the proportion of two embeddings of $v_i$, $\mathbf{\widetilde{z}}_{i_{mid}}$ denotes the further integrated representation of item $v_i$ decided by $\mathbf{z}_{i_{mid}}$.

By the self-attention mechanism, we obtain the further integrated embedding $\mathbf{\widetilde{z}}_{i_{low}}$ and $\mathbf{\widetilde{z}}_{i_{mid}}$ of item $v_i$. Finally, we concat the two embeddings and turn them into a low-dimensional representation:
\begin{equation}
   %\mathbf{h}_{i}=Sum(\mathbf{H'}_{i}^
   \mathbf{\hat{z}}_i=([\mathbf{\widetilde{z}}_{i_{low}} \oplus \mathbf{\widetilde{z}}_{i_{mid}}])\mathbf{W},
   \label{eq:final_eq}
\end{equation}
where $\mathbf{W} \in \mathbb{R}^{2d^{\prime}\times d^{\prime}}$ and $d^{\prime}$ is the embedding size. For item $v_j$, we do the same step to obtain $\mathbf{\hat{z}}_j$.
%Since there are only low and mid-frequency components, the sequence length is 2. The time complexity of both pairwise attention and self-attention mechanisms is $\mathcal{O}(4|d|)$, where $d$ is the embedding dimension. Thus, the overall time complexity of the two-stage attention mechanism is $\mathcal{O}(8|d|)$.
% \begin{gather}
%    \mathbf{\widetilde{z}}_{i_{low}}=\sum_{f}\beta_{i_{low},i_{f}}\mathbf{z}_{i_f}, \quad 
%    \mathbf{\widetilde{z}}_{i_{mid}}=\sum_{f}\beta_{i_{mid},i_{f}}\mathbf{z}_{i_f}, \\
%    \beta_{i_{low},i_f}=\frac{{\mathbf{z}_{j_{low}}}^T\mathbf{z}_{i_f}}{\sum_f{\mathbf{z}_{i_{low}}}^T\mathbf{z}_{i_f}}, \quad 
%    \beta_{i_{mid},i_f}=\frac{{\mathbf{z}_{j_{mid}}}^T\mathbf{z}_{i_f}}{\sum_f{\mathbf{z}_{i_{mid}}}^T\mathbf{z}_{i_f}},
% \end{gather}
% \suhang{simialrly, it's better to separate the above equations into two.} \suhang{explain $\beta$} where $\mathbf{\widetilde{z}}_{i_{low}}$ and $\mathbf{\widetilde{z}}_{i_{mid}}$ denote integrated embedding of item $v_i$ by self-attention mechanism. 

 \subsection{Contrastive Learning Optimization} \label{sec:loss}
 We treat the graph complementary item recommendation as a link prediction problem and follow the principle of contrastive learning \cite{he2020momentum} to construct positive and negative samples for each item, which encourages the model to pull together items that are complementary and push apart those that are not. For each item, positive samples are its complementary items and negative samples are randomly sampled from nodes that do not have links to it. The loss function can be formally defined as :
 \begin{equation}
    \mathcal{L} = -\sum_{e_{i,+}\in\mathcal{E}}\log{\frac{\exp(\mathbf{\hat{z}}_i^T\mathbf{\hat{z}}_+/\tau)}{\sum_{j=0}^{M}\exp(\mathbf{\hat{z}}_i^T\mathbf{\hat{z}}_{j}/\tau)}},
    \label{eq:loss}
 \end{equation}
where $\mathbf{\hat{z}}_+$ is the positive sample, $M$ is the number of the negative items, and $\tau$ is a temperature hyperparameter. Note that, the value of $\mathbf{\hat{z}}_i$ is not fixed, instead, it changes as the item pair changes. In the inference phase, we use the representations generated from the trained model to predict whether two items are complementary. The prediction score can be computed by:
 \begin{equation}
     s_{i,j}=Sigmoid(\mathbf{\hat{z}}_i^T\mathbf{\hat{z}}_j).
     \label{eq:infer}
 \end{equation}
%\suhang{this is for one relationship, can you also write down the final objective function}
%\subsubsection{Training Algorithm} \suhang{add an algorithm and explain the algorithm}

%\subsubsection{Time Complexity} %\suhang{add time complexity analysis}
The algorithm and detailed time complexity analysis can be found in the supplementary files, where the time complexity of spectral-based GCN filters and the two-stage attention mechanism is $\mathcal{O}(3|\mathcal{E}|)$ and $\mathcal{O}(8|d^{\prime}|)$, respectively.

 \section{Experiments}

 In this section, we carry out comprehensive experiments to demonstrate the effectiveness of our method.
 \subsection{Experimental Setup}
 \subsubsection{Datasets}
% \begin{table}[]
% \centering
% \scalebox{0.83}{
% \begin{tabular}{@{}c|ccccc@{}}
% \toprule
% Dataset  & Appliances & Software & Grocery & Toys    & Home    \\ \hline
% \# items    & 4595       & 921      & 100771  & 79577   & 370076  \\
% \# edges    & 22247      & 3460     & 1056902 & 1193698 & 1372435 \\ \bottomrule

% \end{tabular}}
% \caption{Statistics of the datasets.}
% \label{tab:stat}
% \end{table}

Following \cite{liu2020decoupled, hao2020p, bibas2023semi}, we use publicly available benchmark datasets from Amazon. We consider ``also-bought'' as complementary relationships, and our task is to realize the link prediction on the complementary item graphs. We select four datasets: ``Appliances'', ``Grocery'', ``Toys'', and ``Home'', and use the categories and price of each item as features. For categories, we choose BERT \cite{vaswani2017attention} as the pre-trained model to obtain the category embedding, and for the price, we discretize the continuous price to bins using equal-depth binning. Similar to previous work \cite{liu2020decoupled}, for each item, we randomly sample one edge for constructing the test data, one for the validation data, and use the remaining edges as the training data. The statistics of the datasets are shown in Table \ref{tab:s_high}.

 \subsubsection{Baselines and Implementation.}
 The baseline models can be divided into two groups: traditional GNNs and complementary item recommendation models. The first group includes GIN \cite{xu2018powerful} and GraphSage \cite{hamilton2017inductive}. The second group includes Popularity \cite{bibas2023semi}, DCF \cite{galron2018deep}, P-Companion \cite{hao2020p}, and ALCIR \cite{bibas2023semi}. We exclude DecGCN \cite{liu2020decoupled} and EMRIGCN \cite{chen2023enhanced} from our comparison since they incorporate substitutable relationships which are beyond the scope of this paper, as our focus is on complementary item recommendations. For our implementation, we set the embedding size and network layers of both two GCN filters to 16 and 1, respectively. We evaluated the performance using two metrics: Hit Rate (HR@K) and NDCG (Normalized Discounted Cumulative Gain), with $K$ set to 5 and 10. Detailed descriptions of baselines and implementations can be found in the supplementary files and our code is available\footnote{\url{https://github.com/luohaitong/SComGNN}}.

\subsection{Overall Performance}
\begin{table*}
\small
\centering
\scalebox{1}{
\setlength{\tabcolsep}{1.4mm}{
% \begin{tabular}{@{}cc|cc|cc|cc|cc@{}}
% \toprule
% &Dataset & \multicolumn{2}{c}{Appliances}  & \multicolumn{2}{c}{Toys}  & \multicolumn{2}{c}{Grocery} & \multicolumn{2}{c}{Home}    \\ \hline
% & Metric        & HR@10  & NDCG      & HR@10  & NDCG    & HR@10 & NDCG    & HR@10 & NDCG   \\ \hline  \hline
% &Ours     & \textbf{0.7127}      & \textbf{0.4377}              & \textbf{0.8589}      & \textbf{0.5501}      & \textbf{0.8565}   & \textbf{0.5959}     & \textbf{0.8789}    & \textbf{0.6610} \\ \hline
%  & \textit{w/o} l          & 0.1855  & 0.2457       & 0.1365        & 0.2236       & 0.2194        & 0.2560        & 0.1745       & 0.2375        \\
% & \textit{w/o} m       & 0.6420  & 0.4235    & 0.7186        & 0.4907    & 0.7839        & 0.5389    & 0.8387       & 0.6239        \\
%  & \textit{w/o} a       & 0.6766         & 0.4195   & 0.8228        & 0.5149    & 0.8034   & 0.5330       & 0.8626   & 0.6223       \\
% \bottomrule
% \end{tabular}}
\begin{tabular}{@{}cc|ccc|ccc|ccc|ccc@{}}
\toprule
&Dataset & \multicolumn{3}{c}{Appliances}  & \multicolumn{3}{c}{Toys}  & \multicolumn{3}{c}{Grocery} & \multicolumn{3}{c}{Home}    \\ \hline
& Metric      & HR@5  & HR@10  & NDCG    & HR@5  & HR@10  & NDCG   & HR@5 & HR@10 & NDCG   & HR@5 & HR@10 & NDCG   \\ \hline  \hline
&Ours  & \textbf{0.4919}    & \textbf{0.7127}      & \textbf{0.4377}        & \textbf{0.6561}      & \textbf{0.8589}      & \textbf{0.5501}    & \textbf{0.7207}   & \textbf{0.8565}   & \textbf{0.5959}    & \textbf{0.7943}     & \textbf{0.8789}    & \textbf{0.6610} \\ \hline
 & \textit{w/o} l     & 0.1071      & 0.1855  & 0.2457   & 0.0724      & 0.1365        & 0.2236      & 0.1272     & 0.2194        & 0.2560     & 0.0957     & 0.1745       & 0.2375        \\
& \textit{w/o} m     & 0.4364      & 0.6420  & 0.4235   & 0.5194      & 0.7186        & 0.4907      & 0.6113     & 0.7839        & 0.5389     & 0.7349     & 0.8387       & 0.6239        \\
 & \textit{w/o} a     &0.4644        & 0.6766         & 0.4195    & 0.5889    & 0.8228        & 0.5149   & 0.6194    & 0.8034   & 0.5330     & 0.7532     & 0.8626   & 0.6223       \\
\bottomrule
\end{tabular}}}
%\vspace{-8pt}
%\vskip -1em
\caption{The ablation study performance.}
\label{tab:ablation}
%\vspace{-4pt}
\end{table*}
We present our experimental results in Table \ref{tab:overall_perform}, including the results of our model and baselines on four datasets, where the boldfaced and underlined values represent the best and the second-best performance, respectively. Based on the results, we can make the following observations:
% \begin{itemize}
%     \item SComGNN outperforms all other models on all datasets, making it a state-of-the-art model for complementary item recommendation. Specifically, for the HR@10 score, SComGNN outperforms the best baseline model by 7.8\%, 14.3\%, 13.3\%, and 4.4\% on the four datasets, respectively. On the other two metrics(i.e., HR@5 and NDCG), SComGNN achieves a similar improvement in performance. The results demonstrate the importance of leveraging both low-frequency relevance and mid-frequency dissimilarity to enhance the performance of complementary recommendations.
%     \item We observe that traditional graph-based models can also perform well, even without custom modifications for complementary relationships. In fact, some of these models outperform non-graph-based complementary item recommendation models. This highlights the powerful capabilities of GNNs in modeling relationships. Compared to GIN, Graphsage's strong performance comes from the operation on neighbor node sampling, which reduces its reliance on low-frequency components.
% \end{itemize}

First, SComGNN outperforms all other models on all datasets, making it a state-of-the-art model for complementary item recommendation. Specifically, for the HR@10 score, SComGNN outperforms the best baseline model by 7.8\%, 14.3\%, 13.3\%, and 4.4\% on the four datasets, respectively. On the other two metrics (i.e., HR@5 and NDCG), SComGNN achieves a similar improvement in performance. The results demonstrate the importance of leveraging both low-frequency relevance and mid-frequency dissimilarity to enhance performance.

Additionally, we observe that traditional graph-based models can also perform well, even without custom modifications for complementary relationships. In fact, some of these models outperform non-graph-based complementary item recommendation models. This highlights the powerful capabilities of GNNs in modeling relationships. Compared to GIN, Graphsage's strong performance comes from the operation on neighbor node sampling, which reduces its reliance on low-frequency components.

 % First, SComGNN outperforms all other models on all datasets, making it a state-of-the-art model for complementary item recommendation. Specifically, for the HR@10 score, SComGNN outperforms the best baseline model by 7.8\%, 14.3\%, 13.3\%, and 4.4\% on the four datasets, respectively. On the other two metrics(i.e., HR@5 and NDCG), SComGNN achieves a similar improvement in performance. The results demonstrate the importance of leveraging both low-frequency relevance and mid-frequency dissimilarity to enhance the performance of complementary recommendations.

 % Furthermore, we observe that traditional graph-based models can also perform well, even without custom modifications for complementary relationships. In fact, some of these models outperform non-graph-based complementary item recommendation models. This highlights the powerful capabilities of GNNs in modeling relationships. Compared to GIN, Graphsage's strong performance comes from the operation on neighbor node sampling, which reduces its reliance on low-frequency components.
 
 \subsection{Ablation Study}
% \begin{table}
% \small
% \centering
% \caption{The ablation study performance.}
% \vspace{-4pt}
% \scalebox{0.65}{
% \setlength{\tabcolsep}{0.75mm}{
% \begin{tabular}{@{}cc|ccc|ccc|ccc|ccc@{}}
% \toprule
% &Dataset & \multicolumn{3}{c}{Appliances}  & \multicolumn{3}{c}{Toys}  & \multicolumn{3}{c}{Grocery} & \multicolumn{3}{c}{Home}    \\ \hline
% & Metric      & HR@5  & HR@10  & NDCG    & HR@5  & HR@10  & NDCG   & HR@5 & HR@10 & NDCG   & HR@5 & HR@10 & NDCG   \\ \hline  \hline
% &Ours  & \textbf{0.4919}    & \textbf{0.7127}      & \textbf{0.4377}        & \textbf{0.6561}      & \textbf{0.8589}      & \textbf{0.5501}    & \textbf{0.7207}   & \textbf{0.8565}   & \textbf{0.5959}    & \textbf{0.7943}     & \textbf{0.8789}    & \textbf{0.6610} \\ \hline
%  & \textit{w/o} l     & 0.1071      & 0.1855  & 0.2457   & 0.0724      & 0.1365        & 0.2236      & 0.1272     & 0.2194        & 0.2560     & 0.0957     & 0.1745       & 0.2375        \\
% & \textit{w/o} m     & 0.4364      & 0.6420  & 0.4235   & 0.5194      & 0.7186        & 0.4907      & 0.6113     & 0.7839        & 0.5389     & 0.7349     & 0.8387       & 0.6239        \\
%  & \textit{w/o} a     &0.4644        & 0.6766         & 0.4195    & 0.5889    & 0.8228        & 0.5149   & 0.6194    & 0.8034   & 0.5330     & 0.7532     & 0.8626   & 0.6223       \\
% \bottomrule
% \end{tabular}}}
% \label{tab:ablation}
% %\vspace{-8pt}
% \end{table}

 \begin{figure}[t]
    \centering
    \includegraphics[scale=0.255]{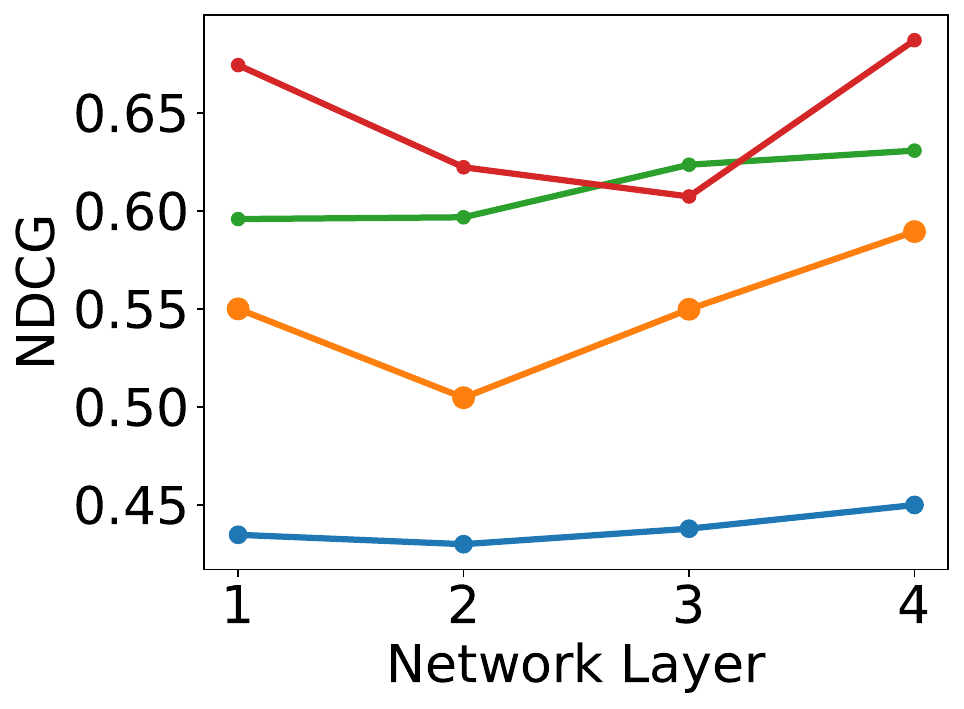}
    \includegraphics[scale=0.255]{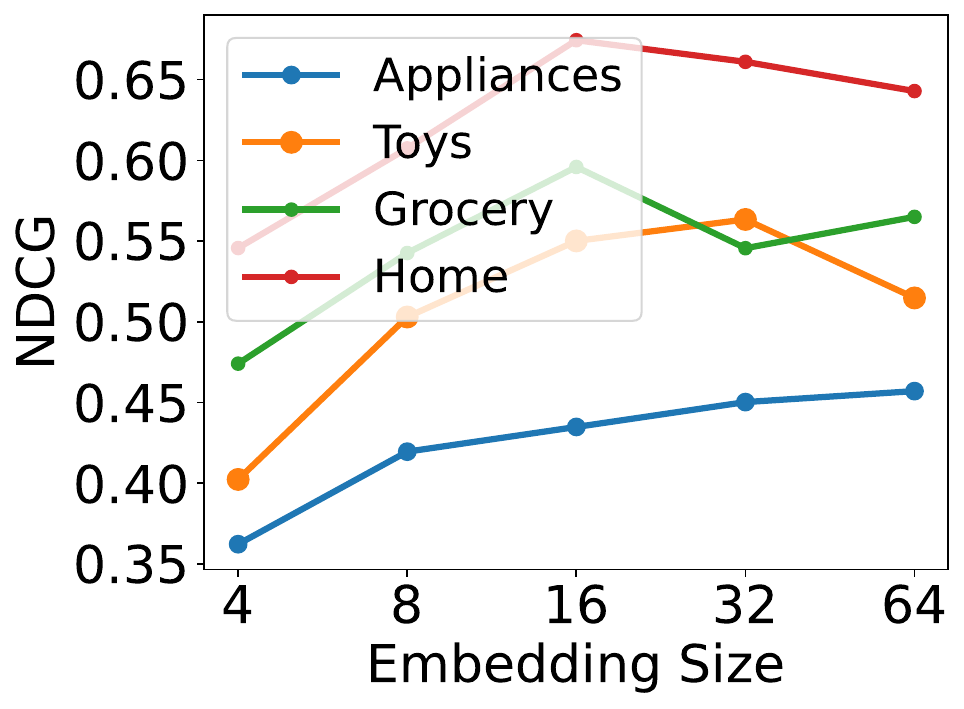}
    %\vspace{-8pt}
    \caption{Hyperparameter sensitivity evaluation.}
    % \vskip -1em
    %\vspace{-10pt}
    \label{fig:sensitivity}
\end{figure}

 We carry out ablation experiments to investigate the contributions of three key modules, i.e., the low-pass GCN filter, the mid-pass GCN filter, and the two-stage attention mechanism. SComGNN \textit{w/o} l is the variant without the low-frequency GCN filter and the two-stage attention mechanism, thus the model only obtains the mid-frequency representation. SComGNN \textit{w/o} m is the variant without the mid-frequency GCN filter and the two-stage attention mechanism, thus the model only obtains the low-frequency representation. SComGNN \textit{w/o} a is the variant without the two-stage attention mechanism thus the mid-frequency and low-frequency representations are simply concatenated. The results are shown in Table \ref{tab:ablation}.
 %Due to similar trends and page limitations, we only show results on two metrics in Table \ref{tab:ablation}, and complete results on three metrics can be found in supplementary files.
 %\suhang{put the ablation study results in another table}

 In Table \ref{tab:ablation}, we can observe that (1) SComGNN achieves the best performance among the four models, indicating the collective importance of all three modules. (2) Compared to SComGNN \textit{w/o} l, ScomGNN \textit{w/o} m achieves better performance, which verifies our observations that the low-frequency component is greater than the mid-frequency component. (3) In some cases, the performance of SComGNN \textit{w/o} a can even be inferior to that of SComGNN \textit{w/o} m. This indicates although the mid-frequency representation is valuable, its integration needs to be more efficient.
 %which highlights the necessity for effective methods of integrating low-frequency and mid-frequency representation.
 
 \subsection{Hyperparameter Analysis}

 We investigate the impact of two hyperparameters on the performance of our model, i.e., the depth of low-pass and mid-pass GCN layers and the embedding size. Due to similar trends and page limitations, we only present the NDCG results in Figure \ref{fig:sensitivity}, and complete results can be found in supplementary files. We can make some conclusions. First, for the depth of GCN layers, the performance may decrease as the depth of network layers increases. This is because our model aggregates structural information from different perspectives so that a one-layer network can perform well. Next, for the embedding size, 16 is the most appropriate value. Increasing the embedding size to 32 and 64 not only does not necessarily improve the model performance but also increases the model complexity and training time. Also, reducing the embedding size to 8 or 4 results in a significant drop in performance, indicating the importance of learning rich and expressive feature representations.

 \subsection{Case Study}
\begin{figure}
    \centering
    \includegraphics[scale=0.2685]{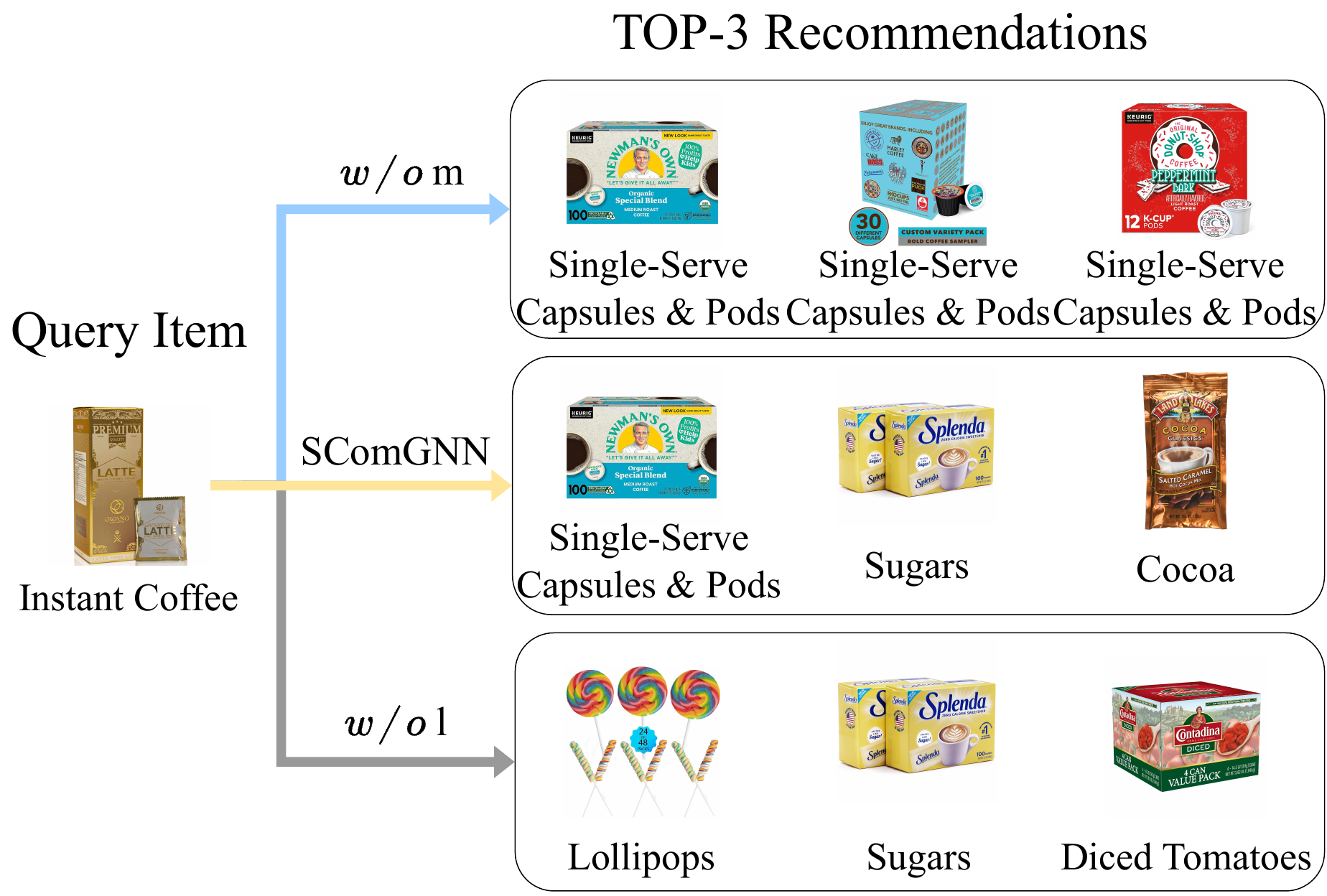}
    %\vspace{-15pt}
    %\vspace{-3pt}
    \caption{Examples of complementary recommendation results with different frequency components.}
    %\vspace{-15pt}
    \label{fig:case_study}
    %\vspace{-10pt}
\end{figure}

To assess the impact of low and mid-frequency components in the production environment, we compare the performance of three models: SComGNN \textit{w/o} m, SComGNN \textit{w/o} l, and SComGNN.
% , which are used to recommend complementary items for a given query item. 
Figure \ref{fig:case_study} shows the TOP-3 complementary items recommended for ``Instant Coffee''. 
%Due to the space limitation, we only show the results on the Grocery dataset, and other datasets show similar results. 
In the recommendations solely based on the low-frequency components, items are all coffee, which is strongly similar to the query item. Conversely, with only the mid-frequency components, items appear to have a low correlation to the query item. However, with both the low-frequency and mid-frequency components, a diverse range of items, including coffee, sugar, and cocoa, are recommended. The inclusion of sugar as a flavoring for coffee, and the presence of cocoa alongside coffee as distinct but related beverages, illustrate our ability to capture both relevance and dissimilarity. This outcome highlights the crucial roles played by the low-frequency component in representing relevance and the mid-frequency component in representing dissimilarity, both of which are essential for complementary relationships.
%As sugar can be used as a flavoring for coffee, and cocoa and coffee are the same beverage but not the same commodity, the recommended items satisfy both relevance and dissimilarity attributes. When we integrate the low-frequency component with the high-frequency component, the recommendations satisfy both relevance and dissimilarity attributes. This result illustrates that the low-frequency component represents the relevance attribute and the mid-frequency component represents the dissimilarity attribute, both of which are essential for complementary relationships.

 \section{Conclusion}
 In this paper, we bridge the gap between spectral properties and attributes of complementary relationships. Our analysis reveals that complementary item graphs primarily consist of low-frequency and mid-frequency components in the spectral domain, representing relevance and dissimilarity attributes. We propose GCN filters to extract the two components and employ a two-stage attention mechanism for adaptive integration. Experiments on four publicly available datasets demonstrate the effectiveness of our theoretical analysis and the proposed method. In the future, more effective GCN filters and integration approaches of two frequency components deserve to be explored.

\section{Acknowledgements}
This work is supported in whole or in part, by National Science Foundation of China (61972381, 62106249, 62372429), Project on Cyber Security and Informatization of Chinese Academy of Sciences (CAS-WX2022SF-0401), and the Pilot for Major Scientific Research Facility of Jiangsu Province of China (NO.BM2021800).

 \bibliography{aaai24}

\begin{thebibliography}{29}
\providecommand{\natexlab}[1]{#1}

\bibitem[{Balcilar et~al.(2020)Balcilar, Renton, H{\'e}roux, Gauzere, Adam, and Honeine}]{balcilar2020bridging}
Balcilar, M.; Renton, G.; H{\'e}roux, P.; Gauzere, B.; Adam, S.; and Honeine, P. 2020.
\newblock Bridging the gap between spectral and spatial domains in graph neural networks.
\newblock \emph{arXiv preprint arXiv:2003.11702}.

\bibitem[{Bibas, Shalom, and Jannach(2023)}]{bibas2023semi}
Bibas, K.; Shalom, O.~S.; and Jannach, D. 2023.
\newblock Semi-supervised Adversarial Learning for Complementary Item Recommendation.
\newblock \emph{arXiv preprint arXiv:2303.05812}.

\bibitem[{Bo et~al.(2021)}]{bo2021beyond}
Bo, D.; et~al. 2021.
\newblock Beyond low-frequency information in graph convolutional networks.
\newblock In \emph{AAAI}, 3950--3957.

\bibitem[{Bruna et~al.(2013)Bruna, Zaremba, Szlam, and LeCun}]{bruna2013spectral}
Bruna, J.; Zaremba, W.; Szlam, A.; and LeCun, Y. 2013.
\newblock Spectral networks and locally connected networks on graphs.
\newblock \emph{arXiv preprint arXiv:1312.6203}.

\bibitem[{Cen et~al.(2019)Cen, Zou, Zhang, Yang, Zhou, and Tang}]{cen2019representation}
Cen, Y.; Zou, X.; Zhang, J.; Yang, H.; Zhou, J.; and Tang, J. 2019.
\newblock Representation learning for attributed multiplex heterogeneous network.
\newblock In \emph{KDD}, 1358--1368.

\bibitem[{Chen et~al.(2023)Chen, He, Xu, Feng, Liu, Song, Yao, and Qiao}]{chen2023enhanced}
Chen, H.; He, J.; Xu, W.; Feng, T.; Liu, M.; Song, T.; Yao, R.; and Qiao, Y. 2023.
\newblock Enhanced Multi-Relationships Integration Graph Convolutional Network for Inferring Substitutable and Complementary Items.
\newblock In \emph{AAAI}, volume~37, 4157--4165.

\bibitem[{Defferrard, Bresson, and Vandergheynst(2016)}]{defferrard2016convolutional}
Defferrard, M.; Bresson, X.; and Vandergheynst, P. 2016.
\newblock Convolutional neural networks on graphs with fast localized spectral filtering.
\newblock \emph{Neurips}, 29.

\bibitem[{Galron et~al.(2018)Galron, Brovman, Chung, Wieja, and Wang}]{galron2018deep}
Galron, D.~A.; Brovman, Y.~M.; Chung, J.; Wieja, M.; and Wang, P. 2018.
\newblock Deep item-based collaborative filtering for sparse implicit feedback.
\newblock \emph{arXiv preprint arXiv:1812.10546}.

\bibitem[{Gao, Wang, and Ji(2018)}]{gao2018large}
Gao, H.; Wang, Z.; and Ji, S. 2018.
\newblock Large-scale learnable graph convolutional networks.
\newblock In \emph{KDD}, 1416--1424.

\bibitem[{Hamilton, Ying, and Leskovec(2017)}]{hamilton2017inductive}
Hamilton, W.; Ying, Z.; and Leskovec, J. 2017.
\newblock Inductive representation learning on large graphs.
\newblock \emph{Neurips}, 30.

\bibitem[{Han et~al.(2007)Han, Cheng, Xin, and Yan}]{han2007frequent}
Han, J.; Cheng, H.; Xin, D.; and Yan, X. 2007.
\newblock Frequent pattern mining: current status and future directions.
\newblock \emph{Data mining and knowledge discovery}, 15(1): 55--86.

\bibitem[{Hao et~al.(2020)Hao, Zhao, Li, Dong, Faloutsos, Sun, and Wang}]{hao2020p}
Hao, J.; Zhao, T.; Li, J.; Dong, X.~L.; Faloutsos, C.; Sun, Y.; and Wang, W. 2020.
\newblock P-companion: A principled framework for diversified complementary product recommendation.
\newblock In \emph{CIKM}, 2517--2524.

\bibitem[{He et~al.(2020)He, Fan, Wu, Xie, and Girshick}]{he2020momentum}
He, K.; Fan, H.; Wu, Y.; Xie, S.; and Girshick, R. 2020.
\newblock Momentum contrast for unsupervised visual representation learning.
\newblock In \emph{CVPR}, 9729--9738.

\bibitem[{He and McAuley(2016)}]{he2016ups}
He, R.; and McAuley, J. 2016.
\newblock Ups and downs: Modeling the visual evolution of fashion trends with one-class collaborative filtering.
\newblock In \emph{WWW}, 507--517.

\bibitem[{Kipf and Welling(2016)}]{kipf2016semi}
Kipf, T.~N.; and Welling, M. 2016.
\newblock Semi-supervised classification with graph convolutional networks.
\newblock \emph{arXiv preprint arXiv:1609.02907}.

\bibitem[{Liu et~al.(2020)Liu, Gu, Ding, Gao, Guo, Bao, and Yan}]{liu2020decoupled}
Liu, Y.; Gu, Y.; Ding, Z.; Gao, J.; Guo, Z.; Bao, Y.; and Yan, W. 2020.
\newblock Decoupled graph convolution network for inferring substitutable and complementary items.
\newblock In \emph{CIKM}, 2621--2628.

\bibitem[{McAuley, Pandey, and Leskovec(2015)}]{mcauley2015inferring}
McAuley, J.; Pandey, R.; and Leskovec, J. 2015.
\newblock Inferring networks of substitutable and complementary products.
\newblock In \emph{KDD}, 785--794.

\bibitem[{Meng et~al.(2018)Meng, Wang, Shu, Li, Chen, Liu, and Zhang}]{MengWSLCLZ18}
Meng, X.; Wang, S.; Shu, K.; Li, J.; Chen, B.; Liu, H.; and Zhang, Y. 2018.
\newblock Personalized Privacy-Preserving Social Recommendation.
\newblock In \emph{AAAI}, 3796--3803.

\bibitem[{Tang et~al.(2022)Tang, Li, Gao, and Li}]{tang2022rethinking}
Tang, J.; Li, J.; Gao, Z.; and Li, J. 2022.
\newblock Rethinking graph neural networks for anomaly detection.
\newblock In \emph{ICML}, 21076--21089.

\bibitem[{Vaswani et~al.(2017)Vaswani, Shazeer, Parmar, Uszkoreit, Jones, Gomez, Kaiser, and Polosukhin}]{vaswani2017attention}
Vaswani, A.; Shazeer, N.; Parmar, N.; Uszkoreit, J.; Jones, L.; Gomez, A.~N.; Kaiser, {\L}.; and Polosukhin, I. 2017.
\newblock Attention is all you need.
\newblock \emph{Neurips}, 30.

\bibitem[{Veli{\v{c}}kovi{\'c} et~al.(2017)Veli{\v{c}}kovi{\'c}, Cucurull, Casanova, Romero, Lio, and Bengio}]{velivckovic2017graph}
Veli{\v{c}}kovi{\'c}, P.; Cucurull, G.; Casanova, A.; Romero, A.; Lio, P.; and Bengio, Y. 2017.
\newblock Graph attention networks.
\newblock \emph{arXiv preprint arXiv:1710.10903}.

\bibitem[{Wang, Sarwar, and Sundaresan(2011)}]{wang2011utilizing}
Wang, J.; Sarwar, B.; and Sundaresan, N. 2011.
\newblock Utilizing related products for post-purchase recommendation in e-commerce.
\newblock In \emph{Proceedings of the fifth ACM conference on Recommender systems}, 329--332.

\bibitem[{Wang et~al.(2018)Wang, Jiang, Ren, Tang, and Yin}]{wang2018path}
Wang, Z.; Jiang, Z.; Ren, Z.; Tang, J.; and Yin, D. 2018.
\newblock A path-constrained framework for discriminating substitutable and complementary products in e-commerce.
\newblock In \emph{WSDM}, 619--627.

\bibitem[{Wu, Zhou, and Zhou(2022)}]{wu2022towards}
Wu, L.; Zhou, Y.; and Zhou, D. 2022.
\newblock Towards high-order complementary recommendation via logical reasoning network.
\newblock In \emph{ICDM}, 1227--1232.

\bibitem[{Wu et~al.(2020)Wu, Pan, Chen, Long, Zhang, and Philip}]{wu2020comprehensive}
Wu, Z.; Pan, S.; Chen, F.; Long, G.; Zhang, C.; and Philip, S.~Y. 2020.
\newblock A comprehensive survey on graph neural networks.
\newblock \emph{IEEE transactions on neural networks and learning systems}, 32(1): 4--24.

\bibitem[{Wu et~al.(2022)Wu, Pan, Long, Jiang, and Zhang}]{wu2022beyond}
Wu, Z.; Pan, S.; Long, G.; Jiang, J.; and Zhang, C. 2022.
\newblock Beyond low-pass filtering: Graph convolutional networks with automatic filtering.
\newblock \emph{IEEE Transactions on Knowledge and Data Engineering}.

\bibitem[{Xu et~al.(2018)Xu, Hu, Leskovec, and Jegelka}]{xu2018powerful}
Xu, K.; Hu, W.; Leskovec, J.; and Jegelka, S. 2018.
\newblock How powerful are graph neural networks?
\newblock \emph{arXiv preprint arXiv:1810.00826}.

\bibitem[{Yao and Harper(2018)}]{yao2018judging}
Yao, Y.; and Harper, F.~M. 2018.
\newblock Judging similarity: a user-centric study of related item recommendations.
\newblock In \emph{RecSys}, 288--296.

\bibitem[{Zhu et~al.(2023)Zhu, Tang, Zhao, and Wang}]{zhu2023you}
Zhu, H.; Tang, X.; Zhao, T.; and Wang, S. 2023.
\newblock You Need to Look Globally: Discovering Representative Topology Structures to Enhance Graph Neural Network.
\newblock In \emph{PAKDD}, 40--52.

\end{thebibliography}
 \clearpage
   \appendix
  \section{Algorithm and Time Complexity Analysis}

\begin{algorithm}[t]
\caption{SComGNN framework}
\label{alg:algorithm}
\textbf{Input}: A complementary item graph $\mathcal{G} = {\{\mathcal{V}, \mathcal{\mathbf{X}}, {\mathcal{E}}}\}$.\\
\textbf{Parameter}: The number of training epochs $N_{epoch}$, the number of layers $L$, and the embedding size $d^{\prime}$.\\
\textbf{Output}: Probability of an edge $e_{i,j}$ for given items $v_i$ and $v_j$.
\begin{algorithmic}[1] %[1] enables line numbers
\FOR{$t = 1, ..., N_{epoch}$}
    \STATE \emph{//Spectral-based GCN filters}.
    \FOR{each layer $h = 0, ..., L$}
        \STATE Obtain $\mathbf{H}_{low}^{l}$ and $\mathbf{H}_{mid}^{l}$ via Eq. (\ref{eq:spectral_filter}).
    \ENDFOR
    \FOR{$e_{i,j}\in\mathcal{E}$}
        \STATE \emph{//Pairwise attention mechanism}.
        \STATE Obtain $\mathbf{z}_{i_{low}}$ and $\mathbf{z}_{i_{mid}}$ via Eq. (\ref{eq:pair_low}) and Eq. (\ref{eq:pair_mid}).
        \STATE Obtain $\mathbf{z}_{j_{low}}$ and $\mathbf{z}_{j_{mid}}$ via Eq. (\ref{eq:pair_low}) via Eq. (\ref{eq:pair_mid}).
        \STATE \emph{//Self-attention mechansim}.
        \STATE Obtain $\mathbf{\widetilde{z}}_{i_{low}}$ and $\mathbf{\widetilde{z}}_{i_{mid}}$ via Eq. (\ref{eq:self_low}) and Eq. (\ref{eq:self_mid}).
        \STATE Obtain $\mathbf{\widetilde{z}}_{j_{low}}$ and $\mathbf{\widetilde{z}}_{j_{mid}}$ via Eq. (\ref{eq:self_low}) and Eq. (\ref{eq:self_mid}).
        \STATE Obtain $\mathbf{\hat{z}}_i$ via Eq. (\ref{eq:final_eq}).
        \STATE Obtain $\mathbf{\hat{z}}_j$ via Eq. (\ref{eq:final_eq}).
    \ENDFOR

    \STATE Minimize loss via Eq. (\ref{eq:loss}).
\ENDFOR
\STATE \textbf{return} $s_{i,j}$ via Eq. (\ref{eq:infer}).
\end{algorithmic}
\end{algorithm}

The algorithm of SComGNN is shown in Algorithm \ref{alg:algorithm}. We first decouple and extract the low-frequency relevance and mid-frequency dissimilarity with low-pass and mid-pass GCN filters. Then, to adaptively integrate the two components, we utilize a pairwise attention mechanism to integrate them in item pairs and then propose a self-attention mechanism to integrate independently. Finally, the model is optimized with the contrastive learning loss function.

 % \section{}
 Based on the Algorithm \ref{alg:algorithm}, we can make the time complexity analysis. For the spectral-based GCN filters, due to the simple implementation in the spatial domain, the computational complexity of the low-pass and mid-pass GCN filters is $\mathcal{O}(|\mathcal{E}|)$ and $\mathcal{O}(2|\mathcal{E}|)$ respectively, where $|\mathcal{E}|$ is the number of edges. Therefore, the total time complexity of spectral-based GCN filters is $\mathcal{O}(3|\mathcal{E}|)$.
 For the two-stage attention mechanism, since there are only low and mid-frequency components, the sequence length is 2. The time complexity of both pairwise attention and self-attention mechanisms is $\mathcal{O}(4|d^{\prime}|)$, where $d^{\prime}$ is the embedding dimension. Thus, the overall time complexity of the two-stage attention mechanism is $\mathcal{O}(8|d^{\prime}|)$.

 \section{Additional Results}
%  \subsection{Results of Ablation Study}
%  \begin{table*}
% \centering
% %\vspace{-4pt}
% \scalebox{1}{
% \setlength{\tabcolsep}{1mm}{
% \begin{tabular}{@{}cc|ccc|ccc|ccc|ccc@{}}
% \toprule
% &Dataset & \multicolumn{3}{c}{Appliances}  & \multicolumn{3}{c}{Toys}  & \multicolumn{3}{c}{Grocery} & \multicolumn{3}{c}{Home}    \\ \hline
% & Metric      & HR@5  & HR@10  & NDCG    & HR@5  & HR@10  & NDCG   & HR@5 & HR@10 & NDCG   & HR@5 & HR@10 & NDCG   \\ \hline  \hline
% &Ours  & \textbf{0.4919}    & \textbf{0.7127}      & \textbf{0.4377}        & \textbf{0.6561}      & \textbf{0.8589}      & \textbf{0.5501}    & \textbf{0.7207}   & \textbf{0.8565}   & \textbf{0.5959}    & \textbf{0.7943}     & \textbf{0.8789}    & \textbf{0.6610} \\ \hline
%  & \textit{w/o} l     & 0.1071      & 0.1855  & 0.2457   & 0.0724      & 0.1365        & 0.2236      & 0.1272     & 0.2194        & 0.2560     & 0.0957     & 0.1745       & 0.2375        \\
% & \textit{w/o} m     & 0.4364      & 0.6420  & 0.4235   & 0.5194      & 0.7186        & 0.4907      & 0.6113     & 0.7839        & 0.5389     & 0.7349     & 0.8387       & 0.6239        \\
%  & \textit{w/o} a     &0.4644        & 0.6766         & 0.4195    & 0.5889    & 0.8228        & 0.5149   & 0.6194    & 0.8034   & 0.5330     & 0.7532     & 0.8626   & 0.6223       \\
% \bottomrule
% \end{tabular}}}
% \caption{The ablation study performance.}
% \label{tab:whole_ablation}
% %\vspace{-8pt}
% \end{table*}

%  The whole results of the ablation study on three metrics are shown in Table \ref{tab:whole_ablation}. 
%  % Results of the HR@5 show similar trends to other metrics.
 
 \subsection{Results of Hyperparameter Analysis}
\begin{figure}[t]
    \centering
    \subfigure[Sensitivity evaluation on NDCG]{
    \includegraphics[scale=0.255]{layer.pdf}
    \includegraphics[scale=0.255]{emb.pdf}
    }
    \subfigure[Sensitivity evaluation on HR@5]{
    \includegraphics[scale=0.255]{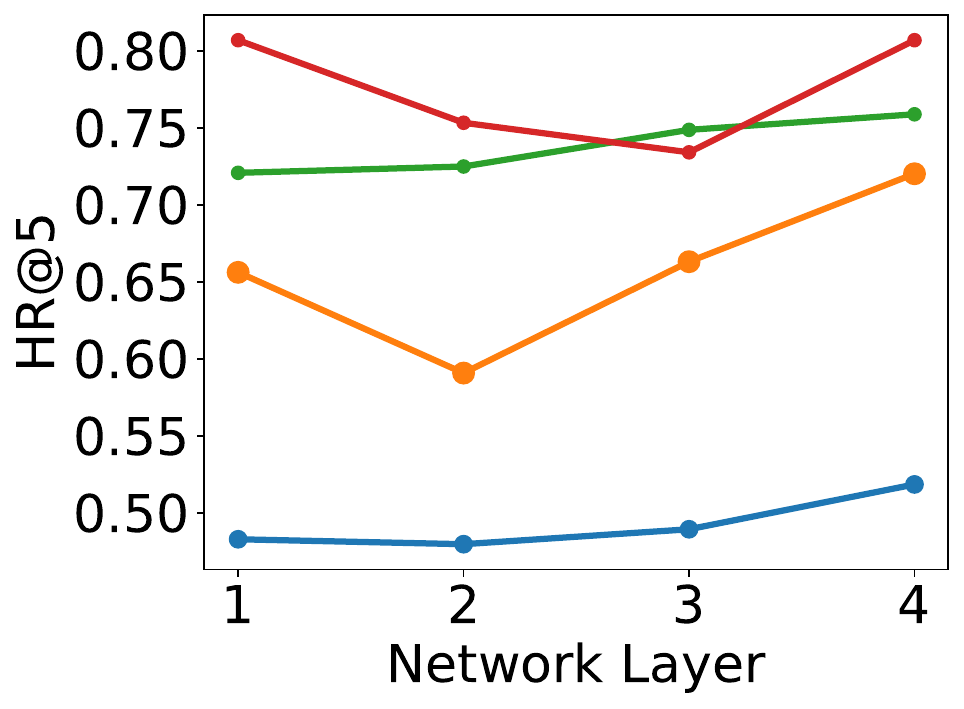}
    \includegraphics[scale=0.255]{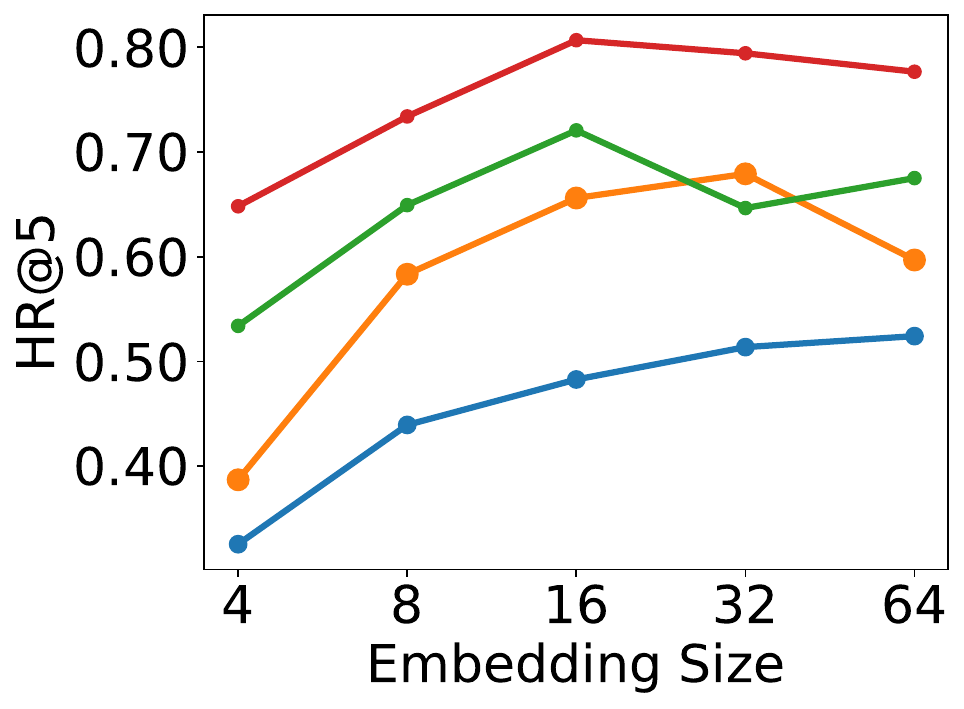}
    }
    \subfigure[Sensitivity evaluation on HR@10]{
    \includegraphics[scale=0.255]{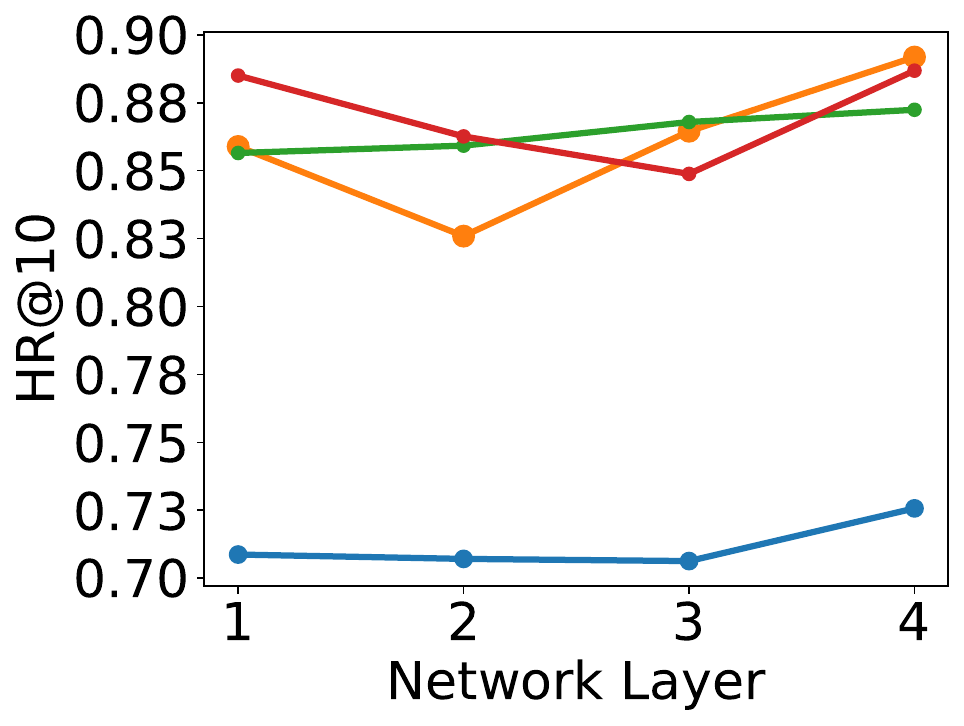}
    \includegraphics[scale=0.255]{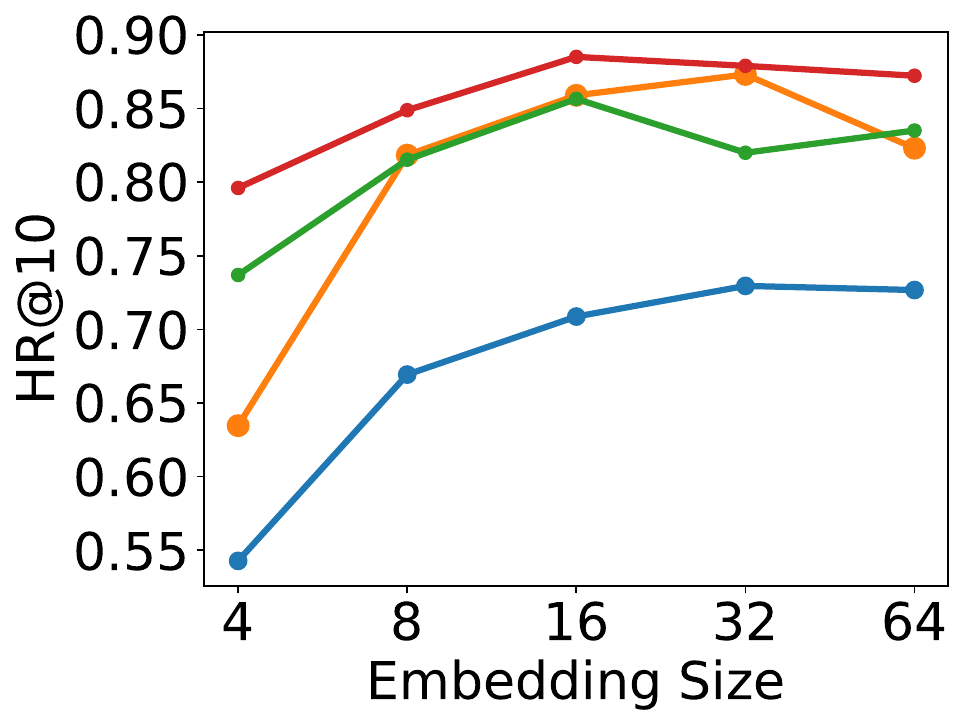}
    }
    \caption{Complete results of hyperparameter sensitivity evaluation.}
    \label{fig:whole_sensit}
\end{figure}

 The whole results of hyperparameter analysis on three metrics are shown in Figure \ref{fig:whole_sensit}. The conclusions drawn from the NDCG score are similar to those drawn from the other metrics.
 
 \section{Baselines and Implementation Details}
 The baseline models can be divided into two groups: traditional GNNs and complementary item recommendation models. The first group includes GIN \cite{xu2018powerful} and GraphSage \cite{hamilton2017inductive}. The second group includes Popularity \cite{bibas2023semi}, DCF \cite{galron2018deep}, P-Companion \cite{hao2020p}, and ALCIR \cite{bibas2023semi}. The descriptions of baseline models are as follows:
   \begin{itemize}
      \item \textbf{GIN}: GIN is a graph neural network that uses a simple and flexible message-passing scheme to learn node embeddings by recursively aggregating information about the local substructure of a graph.
    \item \textbf{GraphSage}:  It is a variant of GCN that aggregates node embeddings from a node's immediate neighbors and a sample of its extended neighborhoods to learn expressive node representation.
    \item \textbf{Popularity}: It is a simple baseline that recommends items based on their popularity. For each item $v_i$, it counts the number of being in complementary item pairs. 
    \item \textbf{DCF}: It is a deep learning-based method focusing on modeling similarity between items.
    \item \textbf{P-Companion}: P-Companion is a deep learning framework that explicitly models both relevance and diversity for complementary item recommendation.
    \item \textbf{ALCIR}: ALCIR is a semi-supervised model to recommend complementary items for seed items.
\end{itemize}
 For Popularity, DCF, P-companion, and ALICR, we refer to the source code provided by \cite{bibas2023semi} \footnote{\url{https://github.com/facebookresearch/cycle_gan_for_complementary_item_recommendations}}. For GIN and GraphSage, we implement them by Pytorch. For a fair comparison, all the baselines follow the same contrastive learning settings and the hyperparameters are tuned on the validation set. We carry out all the experiments with Python 3.7.11, the NVIDIA Tesla V100 GPU, and the 2.60GHz Intel(R) Xeon(R) Gold 6240 CPU. 
\end{document}